\documentclass[DIV=calc, paper=a4, fontsize=11pt, onecolumn]{scrartcl}
\usepackage{amsfonts,amssymb,amsmath}
 \usepackage{lineno,hyperref}
\usepackage{graphicx}
\usepackage{subfig}
\usepackage{float}
\usepackage{tabularx}
\usepackage{authblk} % for author affiliations
\usepackage{arydshln,leftidx,mathtools}
\usepackage{accents}
\usepackage[noend]{algpseudocode}
\modulolinenumbers[1]

\usepackage{xcolor}   

\usepackage[english]{babel}
\usepackage[utf8]{inputenc}
\usepackage{parskip}
\usepackage{multirow, float}
\usepackage{lineno}
\usepackage[top=2.5cm, left=3cm, right=3cm, bottom=4.0cm]{geometry}

 \title{
A mathematical model of \textit{Culex} population abundance and the impact of vector control interventions in a patchy environment
  }
\author[1]{Suman Bhowmick\footnote{Corresponding Author}}
\author[4, 5]{Patrick Irwin}
\author[7]{Kristina Lopez}
\author[6]{Megan Lindsay Fritz}
\author[1, 2, 3]{Rebecca Lee Smith}
\affil[1]{Department of Pathobiology, University of Illinois at Urbana-Champaign, Urbana, Illinois, USA}
\affil[2]{Carl R. Woese Institute for Genomic Biology, University of Illinois at Urbana-Champaign, Urbana, Illinois, USA}
\affil[3]{Carle Illinois College of Medicine, University of Illinois at Urbana-Champaign, Urbana, Illinois, USA}
\affil[4]{Department of Entomology, University of Wisconsin-Madison, Madison, Wisconsin, USA}
\affil[5]{Department of Pathobiological Sciences, University of Wisconsin-Madison, Madison, Wisconsin, USA}
\affil[6]{Department of Entomology, Institute for Advanced Computer Studies, University of Maryland, USA}
\affil[7]{North Shore Mosquito Abatement District, Northfield, Illinois, USA}

\begin{document}
\maketitle
\tableofcontents
\section{Highlights}
\begin{itemize}
\item We have developed a multi-patch weather-driven ordinary differential equation (ODE) mathematical model to describe mosquito population dynamics, incorporating entomological data and weather-dependent variables.
\item  We have calibrated and fit the mosquito trap data from two neighboring sites of Cook county, using weather-driven ODE two patch model.
\item Sensitivity analysis of the \textit{ Basic Offspring Number} within the two-patch framework revealed the mosquito movements between patches are important.
\item We have evaluated the impact of vector control strategies in Cook County under a spatially structured setting and compared the outcomes with those from a single-patch model.
\end{itemize}
\section{Abstract}
Recent mosquito borne outbreaks have highlighted vulnerabilities in our mosquito abatement programmes. 
A possible future outbreak raises a dilemma for mosquito abatement districts to opt for an optimal strategy.
Apparently, spatial dissemination of vector borne disease is caused by the movements of hosts and mosquitoes. 
It should be noted that the vector activity and the spread of mosquito-borne pathogens are entirely intertwined and there is a greater scale of overlap between it.
In our current study, we have developed a mathematical model for the dynamics of Culex mosquito populations in a patchy environment, incorporating entomological data, weather-driven factors, and vector control interventions practiced by the Northwest Mosquito Abatement District (NWMAD), Cook county, Illinois, USA. 
By coupling a temperature-driven multi-patch Ordinary Differential Equation (ODE) model with mosquito abatement strategies utilised by the NWMAD, we have explored how spatial heterogeneity and different mosquito control strategies can potentially affect mosquito abundance. 
We also assess the effectiveness of various vector control strategies, including adulticide and larvicide interventions, under different temporal, spatial configurations.
We further have evaluated how mosquito dispersal influences the effectiveness of various control strategies by comparing single and two-patch model outcomes.
Our simulations demonstrate that models ignoring spatial connectivity via mosquito dispersal can substantially overestimate the efficacy of different intervention strategies or inaccurately represent the threshold levels required for vector persistence.
Through numerical simulations, we have analysed the impact of continuous-pulsatile control measures on population dynamics, providing insight into optimal control strategies for managing \textit{Culex} populations and mitigating the spread of mosquito-borne diseases while considering of mosquito movements in a weather-driven settings in the Cook county, Chicago, Illinois, USA.

\section{Introduction}\label{Introduction}

Mosquitoes of the \textit{Culex} genus complex are among the most widespread and ecologically adaptable vector species globally ~\cite{Gorris1}.  
They play a major role in the transmission of zoonotic and human pathogens, including West Nile virus (WNV), St. Louis encephalitis virus, and avian malaria ~\cite{10.1371/journal.pntd.0006302, doi:10.1128/jvi.01041-24, v13071208} around the globe and especially in North America ~\cite{REISEN2003139, Komar, 10.1093/jmedent/42.3.367, CDC}. 
The abundance and survival of \textit{Culex} mosquitoes are strongly dictated by different environmental factors such as temperature, rainfall, humidity, wind directions, and habitat availability, all of which regulate key life-history traits such as development rates, fecundity, and mortality ~\cite{10.7554/eLife.58511,  10.1093/jmedent/42.1.57,  10.1603/ME11077,  Moser, 10.1603/ME13003, Baril,   10.1093/jme/tjad033}. 
Because of their broad ecological tolerance and capacity to exploit diverse aquatic-terrestrial habitats, the distribution of \textit{Culex} populations are rarely homogeneous.
Instead, their spatial distribution is heterogeneous, and it varies in climate, land use, and resource availability ~\cite{Moser, Gorris1, Fritz, Cardo}. 
Thus, the resulting \textit{Culex} population structure is characterised by the local breeding habitat patches that are connected through adult \textit{Culex} dispersal ~\cite{10.1603/ME11077}. 
Therefore, understanding how environmental heterogeneity and \textit{Culex} mosquito dispersal between the habitat patches together shape the \textit{Culex} population dynamics is essential to predict the relative abundance, and hence for designing effective control strategies ~\cite{10.1603/ME11077, Cardo, 10.1603/ME11272}.
In the era of climate change and global warming, understanding how various climatic factors such as temperature, precipitation, humidity, and wind direction influence \textit{Culex}  population dynamics has become increasingly important ~\cite{su17010102, 10.1093/gbe/evaf143, Samy, Paz, Jordi}. 
Such insights are essential for guiding public health policies and strategies to curb the transmission of mosquito-borne diseases ~\cite{Fed, BHOWMICK2025103163}.\par

Mosquitoes are exothermic organisms, and their life-history traits are highly sensitive to alternations in ambient temperature. 
Recent studies in temperature-dependent mathematical modelling of mosquito populations have enhanced to forecast and to parametrise the mechanistic representation of their life cycles and life history traits.  
Empirical data have established that development, mortality, and fecundity rates follow nonlinear thermal response curves ~\cite{Joac, 10.7554/eLife.58511}
The effects of temperature on various aspects of \textit{Culex} biology have been extensively studied and remain an active area of research ~\cite{Joac, Paz, Fay, Jordi}. 
Temperature not only governs \textit{Culex} population dynamics but also influences the replication and dissemination of WNV within \textit{Culex}, thereby influencing transmission potential ~\cite{Fay, Jordi, Wang,  Maga}. 
Adequate precipitation can provide suitable breeding habitats for \textit{Culex}  by creating egg-laying sites; however, excessive rainfall can have a detrimental effect, washing away eggs and larvae from their habitats, thus yielding an extra mortality ~\cite{Federico, VALDEZ201728, BHOWMICK2025103163, SOH2021142420}. 
Adult \textit{Culex}  mosquitoes also undergo diapause, a process regulated jointly by temperature and photoperiod ~\cite{BHOWMICK2025103163, 10.1603/ME10117,  EZANNO201539}. 
Inclusion of these relationships into stage-structured compartment-based models can allow us to compute the temperature-dependent reproduction metrics such as the Basic Offspring Number ($R_0$) and this metric serves as a threshold for population persistence ~\cite{10.7554/eLife.58511, BHOWMICK2025103163, Joac}. \par

Traditional temperature-driven, compartment-based mosquito population models typically focus on single, isolated patches, assuming environmental homogeneity and often neglecting dispersal amongst the sites~\cite{BHOWMICK2025103163, EZANNO201539, CAILLY20127, ijerph10051698}. 
While these simplifications can help in analytical tractability and simulation, they overlook key ecological processes, especially vector population movements that alone can shape mosquito dynamics in natural landscapes. 
Empirical evidence shows that adult \textit{Culex} can disperse several kilometres in search of hosts or oviposition sites ~\cite{ EZANNO201539, CAILLY20127, BHOWMICK2023110213, Neil}, thus linking otherwise distinct subpopulations into spatially coupled metapopulations. 
Field observations further indicate that neighbouring habitats often differ in temperature profiles and resource availability, leading to asynchronous population dynamics that may become synchronised through migration or external forcing such as coordinated spraying \cite{Felix, Krol, Murdock}. 
These complexities highlight the need for spatially explicit, temperature-driven models capable of quantifying how local environmental heterogeneity and dispersal jointly influence \textit{Culex} population behaviour under different intervention strategies. 
Yet, most existing weather-driven mathematical models rarely include vector dispersal or assess how spatial coupling can alter the efficacy of adulticide or other control measures~\cite{BHOWMICK2023110213, EZANNO201539, CAILLY20127, ijerph10051698}. 
Because movement-driven coupling enables rapid recolonisation of treated habitats and it can potentially undermine any localised interventions.
Mathematical models that ignore dispersal may substantially overestimate control effectiveness or misrepresent persistence thresholds. 
Thus, developing a weather-driven, spatially explicit multi-patch modelling framework can provide a more realistic foundation for understanding how intervention strategies interact with dispersal and environmental gradients to shape \textit{Culex} mosquito abundance.\par

Vector control remains the principal tool for reducing \textit{Culex} populations and interrupting potential vector-borne disease transmission. 
Among the various methods, two chemical interventions are routinely used by different mosquito abatement districts: (i) Larviciding: targets the aquatic immature stages, and (ii) Adulticiding: acts often through ultra-low-volume (ULV) spraying aimed at killing flying adults ~\cite{10.1093/jme/tjad088, BHOWMICK2025103163, 10.1093/jme/tjae041, 10.1371/journal.pone.0332621}.
 These strategies differ significantly in their mechanisms of action, spatial coverage, and duration of impact.
Larvicides act locally by reducing recruitment to the adult population. 
This is done primarily through disrupting the development of larvae into pupae and by inducing additional mortality during the larval stage.
ULV sprays affect  adults across a broader area but typically produce transient effects as the surviving immature stages of \textit{Culex} replenish and the population~\cite{10.1093/jme/tjad088, BHOWMICK2025103163, 10.1093/jme/tjae041}.
In practice, these control strategies are employed with varying frequency and the coordination, ranges from seasonal or periodic treatments to reactive interventions following surveillance triggers ~\cite{10.1093/jme/tjad088, BHOWMICK2025103163, 10.1093/jme/tjae041, 10.1371/journal.pone.0332621, EZANNO201539}.
Despite their widespread use, quantitative understanding of how the timing, frequency, and spatial coordination of these interventions influence \textit{Culex} population dynamics remains limited in the backdrop of weather-driven modelling framework.\par

In our current study we have developed a stage-structured, temperature-driven metapopulation model of \textit{Culex} mosquito population dynamics to investigate how dispersal and vector control interventions shape abundance and persistence across a heterogeneous landscape.
We also have analysed two complementary outcomes: (i) Relative abundance: The temporal evolution of \textit{Culex} abundance in each patch of a two-patch model, and (ii) Magnitude of Basic offspring number ($R_0$): A threshold metric that illustrates whether the \textit{Culex} population will grow or decline in a two-patch model system as a demonstration that can be extended to multipatch modelling framework too. 
We also have compared the effects of single and combined interventions applied homogeneously or heterogeneously across the patches.
Our analysis focuses on two complementary outcomes: (i) the temporal evolution of mosquito abundance in each patch, and (ii) the basic offspring number ($R_0$), a threshold metric that indicates whether the population will grow or decline. 
We compare the effects of single and combined interventions in a single-patch and a two-patch setting.
We aim to identify how dispersal alters and shapes the effectiveness of local interventions utilised by the Northwest Mosquito Abatement District (NWMAD) in the Cook County, Illinois, USA and how spatial coordination can potentially enhance or undermine \textit{Culex} population suppression.

%%%%%%%%%%%%%%%%%%%%%%%%%%%%%%%%%%%%%%%%%%%%%%%%%%%%%%%%%%%%%%%%%%%%%%%%%%%%%%%%%%%%%%%%%%
%%%%%%%%%%%%%%%%%%%%%%%%%%%%%%%%%%%%%%%%%%%%%%%%%%%%%%%%%%%%%%%%%%%%%%%%%%%%%%%%%%%%%%%%%%
%%%%%%%%%%%%%%%%%%%%%%%%%%%%%%%%%%%%%%%%%%%%%%%%%%%%%%%%%%%%%%%%%%%%%%%%%%%%%%%%%%%%%%%%%%

\section{Data collection}\label{Data_Collection}

\subsection{Field site description}\label{Field_site_description}

\begin{figure}[H]
\centering
\includegraphics[width=15cm]{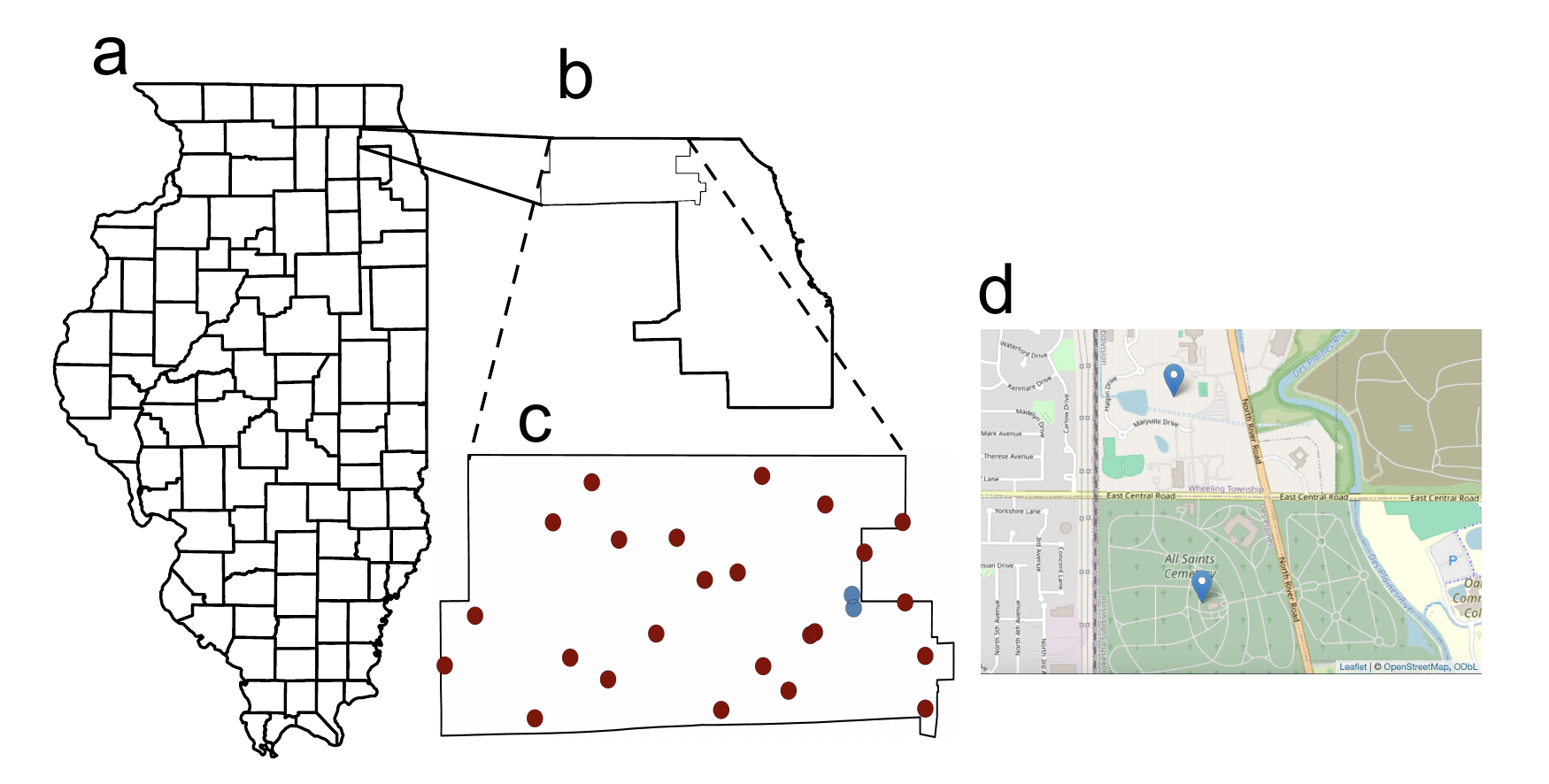}
	\caption{ \textit{Culex} population model \eqref{Eq1}, \eqref{Eq2} and \eqref{Eq3} study area, displaying the locations of mosquito trap data.
	Figure (c) shows the locations of the trap data, represented by the red dots under the jurisdiction of NWMAD within the Cook county (Figure (b)), Illinois, USA (Figure (a))
	 serviced between $2014$ and $2019$. 
	 Two blue dots represent neighbouring trap locations (Figure (c)) and Figure (d) shows the street map features of these two adjacent sites with $7$ traps in each site.
           }
    \label{fig:TrapData}
\end{figure}

The study sites were located within the NWMAD in Cook County, Illinois, USA, an area of approximately $605$ $\text{km}^2$ encompassing the northwest suburbs of Chicago. 
The NWMAD consistently collected and maintained mosquito trapping and identification data throughout the study period \cite{BHOWMICK2024107346, BHOWMICK2025103163}. 
Traps were operated during the summer  from May to September and were collected five days a week (from Monday to Friday), with occasional exceptions for holidays or severe weather. 
The red dots in the Figure\ref{fig:TrapData}(c) indicate the trap geocoordinates, with all sites situated within a $4$ km radius under the jurisdiction of NWMAD. 
For model validation, two nonresidential cemetery sites in Des Plaines, Cook County, Illinois-both located within the jurisdiction of the NWMAD are selected for this study (Figure \ref{fig:TrapData}(d)) to evaluate the performance of the metapopulation model described in the section \ref{Metapopulation_model}.
Each site spanned roughly $0.4-0.8 \text{km}^{²}$ and was paired geographically, with one site in each pair chosen as control and the other as treatment during the same season. 
All sites are located within a $4$ km radius.
Historically, these areas received fewer than one adulticide application per year, typically Zenivex® E20 or Anvil® 10+10. 
All stormwater catch basins and above-ground breeding habitats were treated with $150$ day Altosid® XRT briquets, with follow-up inspections ensuring larvicide effectiveness. 
The study ran for $10$ weeks in $2018$, from June to August (epidemiological weeks $24-33$).
A detailed description about these two trap locations are included in \cite{10.1093/jme/tjad088, 10.1093/jme/tjae041}.

\subsection{Mosquito Collection}\label{Mosquito_collection}
To assess the impact of ULV (adulticide) application on mosquito populations, a total of $28$ mosquito traps were deployed. 
Each site was equipped with CDC gravid traps (Model $1712$, John W. Hock Company, Gainesville, FL), $5$ CDC miniature light trap (Model 512, John W. Hock Company, Gainesville, FL), and BG-Sentinel trap (Biogents, Regensburg, Germany). 
These traps were operated continuously, with collections made daily from Monday to Friday.
CDC light traps and BG-Sentinel traps were baited with carbon dioxide. 
Mosquito surveillance was conducted from epidemiological week $24$ through week $33$. 
All adult mosquitoes were identified utilising morphological characteristics (Siverly $1972$) and subsequently counted. 
\textit{Culex pipiens} and \textit{Culex restuans} specimens collected from BG-Sentinel and CDC light traps were not distinguished and are collectively referred to hereafter as target \textit{Culex} vector species for WNV.

\subsection{Adulticide application}\label{Adulticide_application}
At the treatment sites, five consecutive weekly adulticide were sprayed during epidemiological weeks 26 to 30 whereas control sites received a single spray application during the study period, consistent with district policy.
Zenivex® E20 (Central Life Sciences, Schaumburg, IL) or Anvil 10+10 (Clarke, St. Charles, IL), diluted 1:1 with mineral oil to produce a $10\%$ etofenprox solution, was utilised via ULV spraying using a truck-mounted London Fog unit (London Foggers, Minneapolis, MN). 
ULV spraying occurred no earlier than 30 minutes before sunset to coincide with stable atmospheric conditions. 
ULV applications at nearby treatment sites did not result in drift into the control site, as prevailing winds did not blow in its direction.
Comprehensive descriptions of ULV spray application procedures are provided in \cite{10.1093/jme/tjad088, 10.1093/jme/tjae041}.

\subsection{Weather data}\label{Weatherdata}
\begin{figure}[H]
\centering
\subfloat[Mean daily temperature data ]{\includegraphics[width=0.45\textwidth]{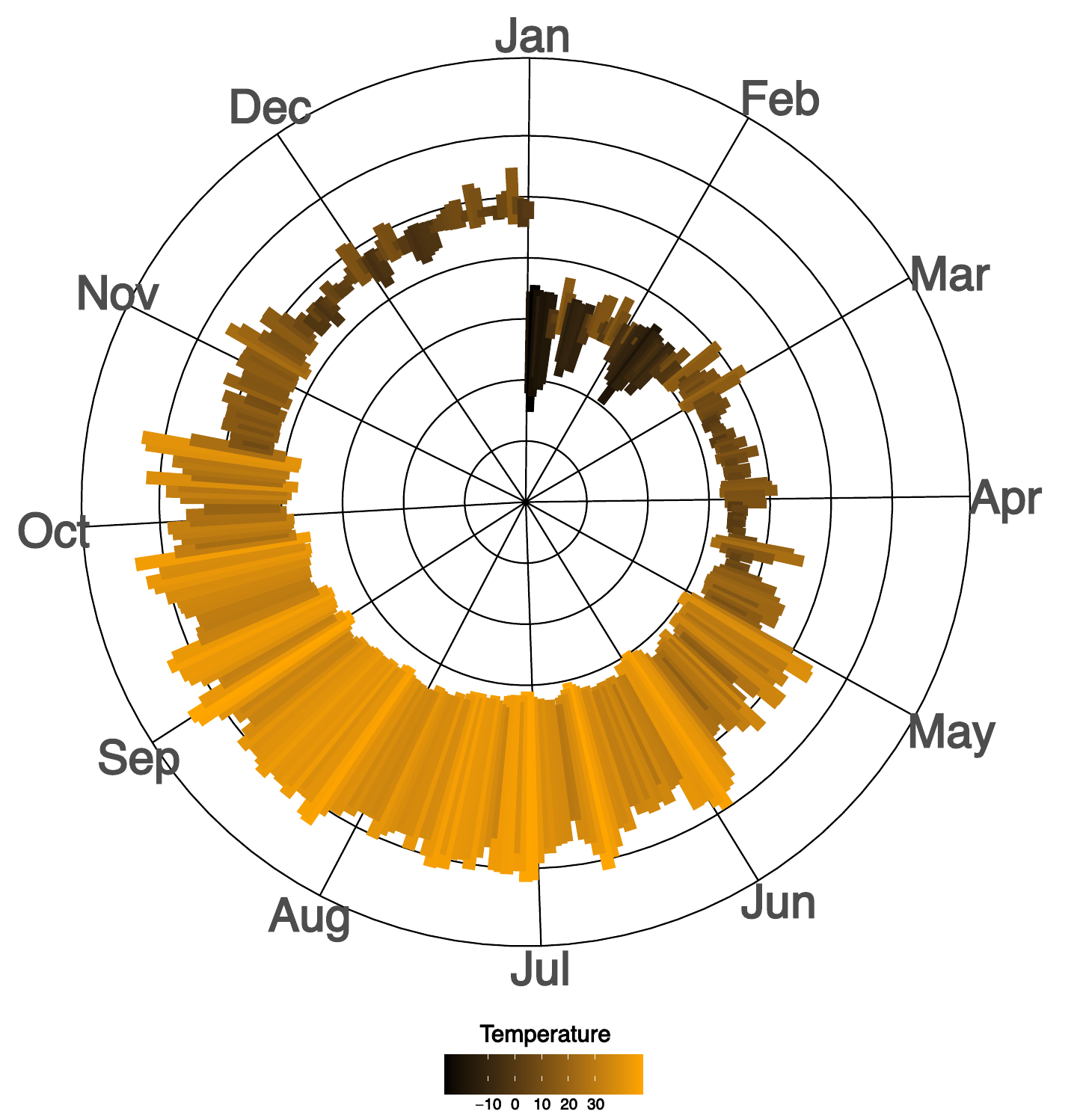 }\label{fig:Temp1}}
\hfill
\subfloat[Precipitation data ]{\includegraphics[width=0.47\textwidth]{  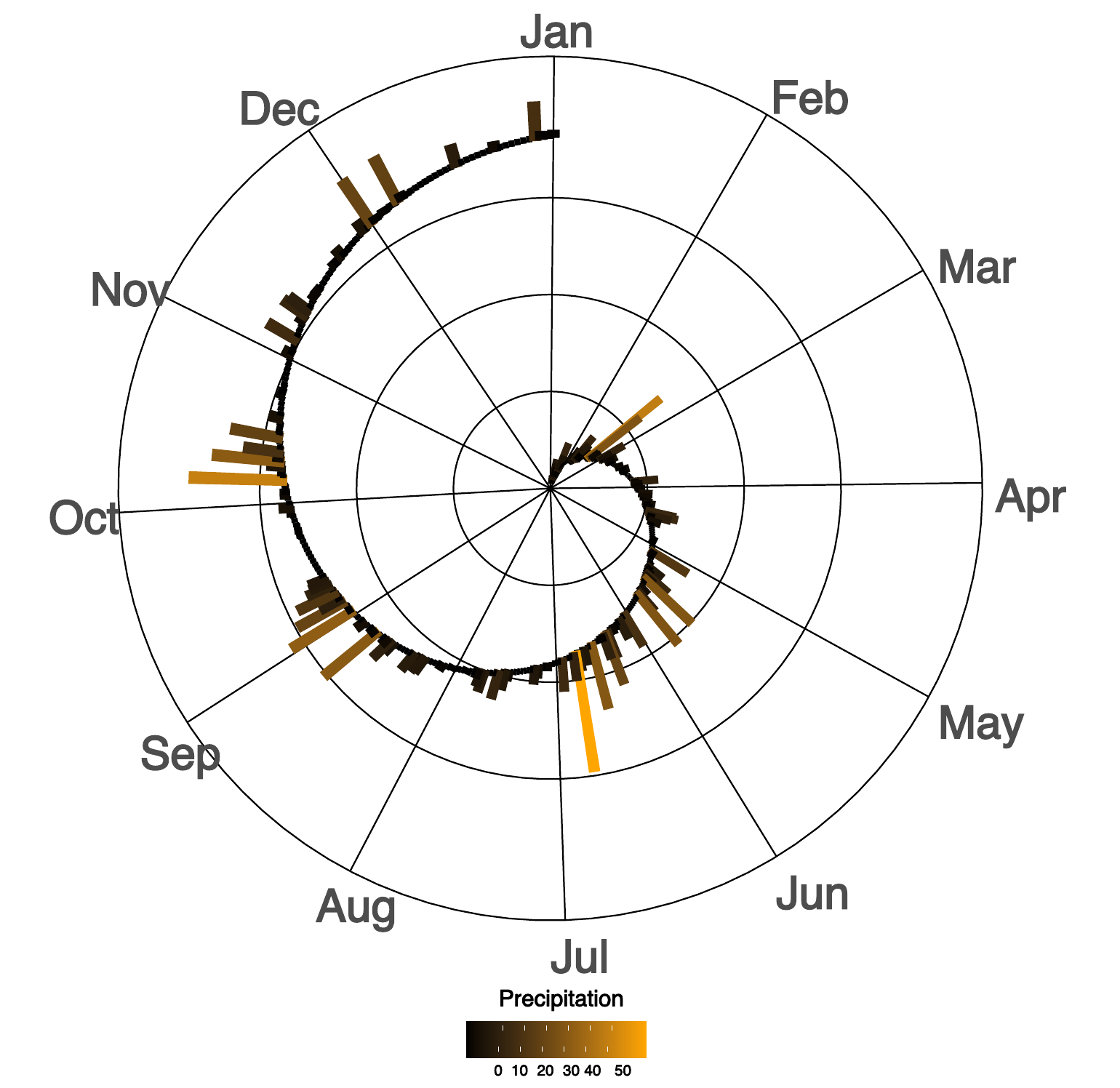 }\label{fig:Prec1}}
\caption{
Time series of daily mean temperature and precipitation data for NWMAD County of $2018$, obtained from the PRISM Climate Group at Oregon State University, which provides gridded estimates at a $4$ km spatial resolution ~\cite{PRISMClimateGroup}.
}\label{fig:WeatherData}
\end{figure}

We have obtained daily mean temperature and precipitation data for $2000–2021$ from the PRISM Climate Group ~\cite{PRISMClimateGroup}. 
Figure~\ref{fig:WeatherData} shows the corresponding time series. 
The PRISM dataset provides $4$ km resolution grids generated by integrating weather station data with topographic information.

%%%%%%%%%%%%%%%%%%%%%%%%%%%%%%%%%%%%%%%%%%%%%%%%%%%%%%%%%%%%%%%%%%%%%%%%%%%%%%%%%%%%%%%%%%
%%%%%%%%%%%%%%%%%%%%%%%%%%%%%%%%%%%%%%%%%%%%%%%%%%%%%%%%%%%%%%%%%%%%%%%%%%%%%%%%%%%%%%%%%%
%%%%%%%%%%%%%%%%%%%%%%%%%%%%%%%%%%%%%%%%%%%%%%%%%%%%%%%%%%%%%%%%%%%%%%%%%%%%%%%%%%%%%%%%%%

\section{Model formulation}\label{Model_formulation}
\subsection{\textit{Culex} life cycle}\label{Culex_life_cycle}
We have developed this model based on the following biological and entomological assumptions. 
We have simplified the life cycle of \textit{Culex} mosquito and comprises two main stages: the aquatic stage (egg, larva, and pupa) and the adult stage as illustrated in Figure~\ref{fig:f2} 
~\cite{doi:10.1086/374473, Moser}.

As soon as they emerge from the final aquatic stage, male and female adults typically mate without delay.
Once inseminated, female \textit{Culex} mosquitoes disperse in search of a blood-feeding host. 
This dispersal can involve long-distance movements, which carry the risks of predation and host defensive behaviours ~\cite{EZANNO201539}.
After obtaining a blood meal, female \textit{Culex} mosquitoes usually retreat to sheltered locations, where they rest for several days while their eggs mature.
During this period, they typically indulge in short, localised movements between resting sites as they are less risky than long-distance dispersal.
Once eggs have matured, female mosquitoes seek suitable aquatic oviposition sites and this once  again involves long-range, high-risk movements. 

\subsection{Single-patch mosquito dynamics}\label{Single_patch_dynamics}
\begin{figure}[H]
\centering
\subfloat[Metapopulation]{\includegraphics[width=0.4\textwidth]{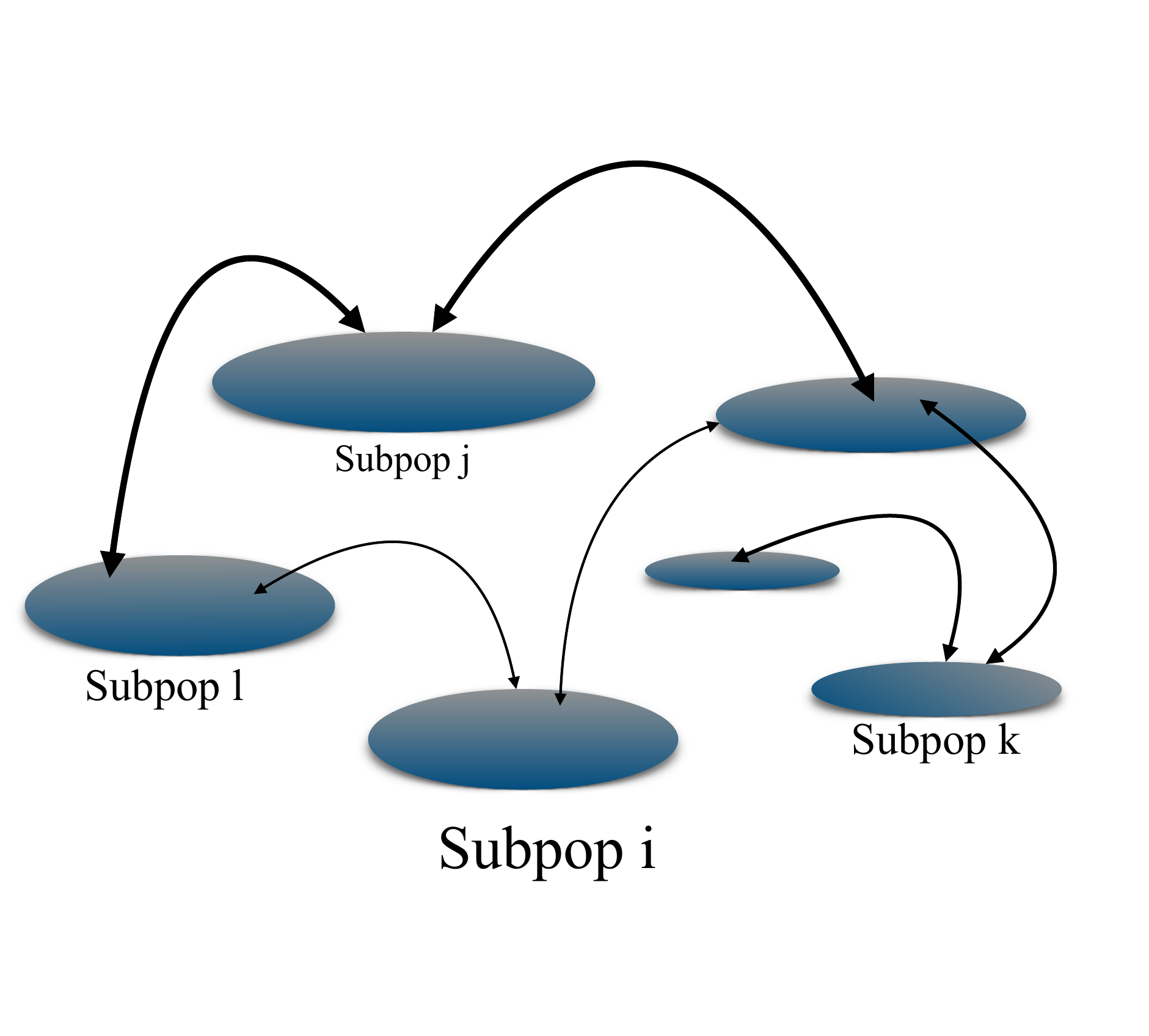}\label{fig:f1}}
\hfill
\subfloat[Simplified lifecycle]{\includegraphics[width=0.4\textwidth]{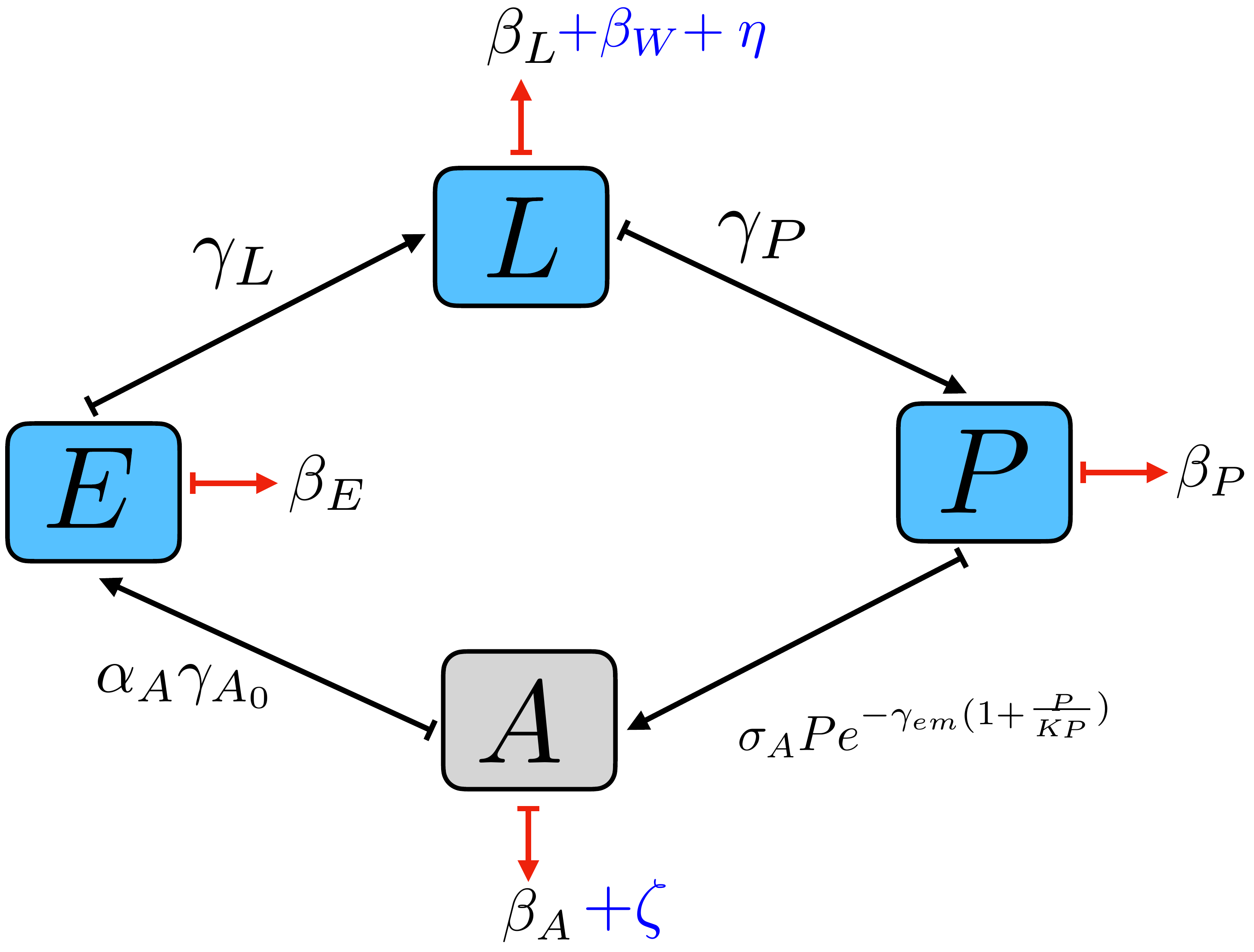}\label{fig:f2}}
\caption{
(a) Schematic illustration of a metapopulation model and the associated movements amongst patches of adult mosquitoes.
It represents different patches in a metapopulation weather-driven model and the connections linking them.
It describes movement of adult individuals between patches in a compartmental model, where each patch hosts a population structured by distinct life stages.
(b) Model diagram of \textit{Culex} population dynamics in temperate climate. 
Aquatic stages are drawn in blue, adult females in grey.
We depict an additional mortality rate due to flushing in larvae population in red ($\beta_W$) and additional death due to larvicide ($\eta$), an extra death due to the application of adulticide in blue ($\zeta$) and as described in  \eqref{Eq1}, \eqref{Eq2} and \eqref{Eq3}.
}\label{fig:ModelFlowMeta}
\end{figure}

We have constructed a generic \textit{Culex} population model following the approaches of ~\cite{CAILLY20127, EZANNO201539, YU201828}. 
The model represents four life stages: three aquatic stages (eggs $E$, larvae $L$, pupae $P$) and the adult female stage ($A$) (Figure~\ref{fig:f2}). 
Stage transitions and maturations are driven by weather-dependent processes that include egg hatching, larval and pupal mortality, pupation, adult emergence, mortality, and oviposition 
~\cite{CAILLY20127, EZANNO201539, YU201828, ijerph10051698, LAPERRIERE201199, BHOWMICK2020110117, BHOWMICK2024107346}. 
We have incorporated density-dependent mortality at the larval stage, and adult emergence success is assumed to depend on pupal density. 
Male mosquitoes are not explicitly modeled.

\begin{eqnarray}\label{Eq1}
 \frac{dE}{dt} &=&  \gamma_{A_{0}}\alpha_A A -  \psi\gamma_L E  -  \beta_E E\\ \nonumber
\frac{dL}{dt} &=&  \psi\gamma_L E-  \beta_LL-  \beta_1 \frac{L^2}{ K_L}-  \gamma_PL- \beta_WL\\ \nonumber
\frac{dP}{dt}&=&  \gamma_{P}L- \beta_P P - \gamma_A P\\ \nonumber
\frac{dA}{dt} &=&  \gamma_A\sigma_A Pe^{-\gamma_{em}(1+\frac{P}{ K_P}})-\beta_AA  \\ \nonumber
\end{eqnarray}
with 
\[
    \psi= 
\begin{cases}
    0,& \text{During diapause} \\
    1,& \text{otherwise}
\end{cases}
\]

In our model, we have represented diapause  as an overwintering stage during which adult \textit{Culex} remain inactive ~\cite{BHOWMICK2020110117, LAPERRIERE201199}. 
Entry into diapause is modelled by a Boolean variable ($\psi$) that defines the start and end of a fixed overwintering period ~\cite{CAILLY20127, EZANNO201539, YU201828}. 
This abstraction aligns with the  field observations that diapause in \textit{Culex} is primarily triggered by photoperiod, modulated by temperature, and occurs after adult emergence and mating but prior to blood-feeding ~\cite{clements2023biology, YU201828}. 
In Cook County, Illinois, we have assumed a long diapause period consistent with extended cold seasons.

\subsection{Metapopulation model}\label{Metapopulation_model}
Our modeling framework is based on a heterogeneous metapopulation network, where each node (habitat patch) represents a local \textit{Culex} subpopulation ~\cite{Neil}. 
Connectivity between patches is established through adult mosquito dispersal along defined migration pathways. 
The local \textit{Culex} population dynamics within each patch are represented by a compartmental structure \eqref{Eq1}, while the overall system is governed by a set of weather-driven coupled ODEs (\eqref{Eq2} and \eqref{Eq3}) that capture both intra-patch dynamics and inter-patch dispersal processes across the network. 
This formulation is ecologically motivated, as the limited flight range of \textit{Culex} mosquitoes restricts their dispersal to nearby habitat patches, thereby shaping the spatial structure of mosquito populations and influencing the potential for pathogen spread.

\subsubsection{Multipatch mosquito dynamics} \label{Multipatch_mosquito_dynamics}

The movement rate of \textit{Culex} from the habitat patch $j$ to the patch $i$ is denoted by $r_{ij}$ ~\cite{Neil, BHOWMICK2023110213}. 
Let $P_i(t)$ represent the population in patch $i$. 
The dynamics of $P_i(t)$ for $i = 1, \dots, n$ can then be expressed as in equation~\eqref{eq:Pi}, 
where $\Pi_{P_i}$ denotes recruitment into the population of patch $i$, 
and $\mu_{P_i}$ represents the mortality rate in patch $i$:  

\begin{equation}
\frac{dP_i}{dt} = \Pi_{P_i} - \mu_{P_i} P_i 
+ \sum_{\substack{j=1 \\ j \neq i}}^n r_{ij} P_j 
- \sum_{\substack{j=1 \\ j \neq i}}^n r_{ji} P_i ,
\label{eq:Pi}
\end{equation}

In a more compact vector--matrix form, the system for all $n$ patches can be written as  

\begin{equation}
\frac{d\mathbf{P}}{dt} = \mathbf{\Pi}_P - \mathbf{M}_P \mathbf{P} + R \mathbf{P},
\label{eq:matrixform}
\end{equation}

where  
\begin{align*}
\mathbf{P} &= 
\begin{bmatrix}
P_1  P_2  \cdots  P_n
\end{bmatrix}^T, 
&
\mathbf{\Pi}_P &= 
\begin{bmatrix}
\Pi_{P_1}  \Pi_{P_2}  \cdots  \Pi_{P_n}
\end{bmatrix}^T, 
&
\mathbf{M}_P &= \text{diag}[\mu_{P_i}]
\\[1em]
R &= [r_{ij}], \quad \text{with } r_{ii} = -\sum_{j \neq i} r_{ji}.
\end{align*}

This formulation provides a compact representation of local demographic processes  and inter-patch dispersal across the metapopulation network.

\subsubsection{Multipatch mosquito dynamics without larvicide and adulticide}\label{Multipatch_mosquito_dynamics_without_larvicide_and_adulticide}

\begin{eqnarray}\label{Eq2}
 \frac{dE_i}{dt} &=&  \mathbf{\gamma_{A_{0}}\alpha_A }A_i -  \mathbf{ \psi\gamma_L} E_i  -   \mathbf{\beta_E} E_i\\ \nonumber
\frac{dL_i}{dt} &=&  \mathbf{\psi\gamma_L }E_i-  \mathbf{\beta_L} L_i-  \mathbf{\beta_1} \frac{L_{i}^2}{ \mathbf{K_L}}-  \mathbf{\gamma_P}L_i- \mathbf{\beta_W} L_i\\ \nonumber
\frac{dP_i}{dt}&=&  \mathbf{\gamma_{P}}L_i- \mathbf{\beta_P} P_i - \mathbf{\gamma_A} P_i\\ \nonumber
\frac{dA_i}{dt} &=&  \mathbf{\gamma_A}\sigma_A P_ie^{-\gamma_{em}(1+\frac{P_i}{ \mathbf{K_P}})}-\mathbf{\beta_A} A_i+ \sum^{n}_{j\neq i}r_{ij}A_j- \sum^{n}_{j\neq i}r_{ji}A_i    \\ \nonumber
\end{eqnarray}

\subsubsection{Multipatch mosquito dynamics with larvicide and adulticide}\label{Multipatch_mosquito_dynamics_with_larvicide_and_adulticide}
With larvicide:
\begin{eqnarray}\label{Eq3}
 \frac{dE_i}{dt} &=& \mathbf{\gamma_{A_{0}}}\mathbf{\alpha_A} A_i -  \psi \mathbf{\gamma_L} E_i  -  \mathbf{\beta_E} E_i\\ \nonumber
\frac{dL_i}{dt} &=& \psi\mathbf{\gamma_L} E_i- \mathbf{\beta_L} L_i- \mathbf{\beta_1} \frac{L_{i}^2}{K_L}- \mathbf{\gamma_P}L_i-\mathbf{\beta_W} L_i-\mathbf{\eta_L} L_i\\ \nonumber
\frac{dP_i}{dt}&=& \mathbf{\gamma_{P}}L_i-\mathbf{\beta_P} P_i -\mathbf{\gamma_A} P_i\\ \nonumber
\frac{dA_i}{dt} &=& \mathbf{\gamma_A}\sigma_A P_ie^{-\gamma_{em}(1+\frac{P_i}{K_P})}-( \mathbf{\beta_A}+ \mathbf{\zeta}) A_i+ \sum^{n}_{j\neq i} \mathbf{r_{ij}} A_j- \sum^{n}_{j\neq i} \mathbf{r_{ji}} A_i    \\ \nonumber
\end{eqnarray}

\subsubsection{Dispersal network}
%Unlike human movement, it is unrealistic to track the flight ways of mosquitoes by specific origins and destinations. 
\textit{Culex} mosquitoes have a limited flight range, and individuals in habitat patch $i$ can potentially interact with neighboring patch $j$ depending on proximity ~\cite{Neil, BHOWMICK2023110213}. 
Given our daily time scale and fine spatial resolution to solve the metapopulation model (\eqref{Eq2}, \eqref{Eq3}), we explicitly incorporate mosquito mobility when modelling abatement strategies. 
We represent this activity using a biological interaction radius around breeding sites: letting $d_{ij}$ denote the distance between habitat patch $i$ and habitat patch $j$, and $d_{\text{max}}$ the maximum interaction radius, we have assumed that mosquitoes from habitat patch $i$ interact with the population of habitat patch $j$ according to a linearly decreasing function of distance ~\cite{Neil, BHOWMICK2023110213} and it is given by:
\begin{equation}
r(d_{ij}) = \begin{cases}
\frac{d_{max}-d_{ij}}{d_{\text{max}}}, ,& d_{ij}<d_{\text{max}}.\\
0, & \text{otherwise}.
\end{cases}
\end{equation}

%%%%%%%%%%%%%%%%%%%%%%%%%%%%%%%%%%%%%%%%%%%%%%%%%%%%%%%%%%%%%%%%%%%%%%%%%%%%%%%%%%%%%%%%%%
%%%%%%%%%%%%%%%%%%%%%%%%%%%%%%%%%%%%%%%%%%%%%%%%%%%%%%%%%%%%%%%%%%%%%%%%%%%%%%%%%%%%%%%%%%
%%%%%%%%%%%%%%%%%%%%%%%%%%%%%%%%%%%%%%%%%%%%%%%%%%%%%%%%%%%%%%%%%%%%%%%%%%%%%%%%%%%%%%%%%%

\begin{table}[H]
\centering
\begin{tabular}{||c c  ||} 
 \hline
 Variables & Definition  \\ [0.5ex] 
 \hline\hline
$E$ & Egg density   \\ 
$L$ & Larvae   density\\
$P$ & Pupae density \\
$A$ & Adult mosquito density \\ 
[1ex] 
 \hline
\end{tabular}
\caption{Model variables and their definitions as mentioned in \eqref{Eq1} in the Cook county, Illinois, USA.}
\label{table:A}
\end{table}

%%%%%%%%%%%%%%%%%%%%%%%%%%%%%%%%%%%%%%%%%%%%%%%%%%%%%%%%%%%%%%%%%%%%%%%%%%%%%%%%%%%%%%%%%%
%%%%%%%%%%%%%%%%%%%%%%%%%%%%%%%%%%%%%%%%%%%%%%%%%%%%%%%%%%%%%%%%%%%%%%%%%%%%%%%%%%%%%%%%%%
%%%%%%%%%%%%%%%%%%%%%%%%%%%%%%%%%%%%%%%%%%%%%%%%%%%%%%%%%%%%%%%%%%%%%%%%%%%%%%%%%%%%%%%%%%
%%%%%%%%%%%%%%%%%%%%%%%%%%%%%%%%%%%%
\begin{table}[H]
\centering
\begin{tabular}{||c c c c||} 
 \hline
 Parameters & Definition   & Values  & Source\\ [0.5ex] 
 \hline\hline
 %% Egg Parameters 
$\alpha_A$    		& Number of eggs laid by egg laying mosquitoes		 		& $350$			&	~\cite{CAILLY20127}					\\ 
$\beta_E$     		& Natural mortality rate of egg 						 		& $0.0262$		&	~\cite{CAILLY20127, EZANNO201539}					\\
$\gamma_L$ 		& Developmental rate of eggs into larvae 				 		& $f(T)$			&	~\cite{CAILLY20127}					\\
%% Larvae Parameters 
$\gamma_P$ 		& Developmental rate of larvae into pupae 					& $f(T)$			&	~\cite{CAILLY20127, EZANNO201539}					\\
$\beta_L$      		& Minimum mortality rate of larvae 							& $0.0304$		&	~\cite{CAILLY20127, EZANNO201539}					\\
$\beta_1$     		& Density dependent mortality rate of larvae 					& $f(T)$			&	~\cite{CAILLY20127, EZANNO201539}					\\
$\beta_W$     		& Mortality rate of larvae due to flushing  						& $0.003$			&	Assumed					\\
$K_L$       	& Standard environment carrying capacity for larvae		 		         & $8\times10^{8}$	&	 ~\cite{CAILLY20127, EZANNO201539, ijerph10051698}		\\
%% Pupae Parameters 
$\gamma_A$ 		& Developmental rate of pupae into adult mosquito 				& $f(T)$			&	~\cite{CAILLY20127}					\\
$\beta_P$     		& Minimum mortality rate of pupae 							& $0.0146$		&	~\cite{CAILLY20127, EZANNO201539, YU201828} 					\\
$K_P$      		& Standard environment carrying capacity for pupae 				& $1\times10^{7}$	&	~\cite{CAILLY20127, EZANNO201539}					\\
%% Adult mosquito parameters 
$\beta_A$     		& Natural mortality rate of adult mosquito 						& $f(T)$			&	~\cite{BHOWMICK2020110117, LAPERRIERE201199}					\\
%$\gamma_B$ 		& Developmental rate of adult mosquito into blood seeking 		& $2$			&	~\cite{CAILLY20127, EZANNO201539}					\\
%$\mu_B$      		& Mortality rate of adult mosquito related to seeking behaviour 		& $0.8$			&	~\cite{CAILLY20127, EZANNO201539}			 		\\ 
%$\gamma_{En}$ 	& Developmental rate of blood seeking mosquito into engorged 	& $2$			&	~\cite{CAILLY20127, EZANNO201539}					\\
%$\gamma_{El}$  	& Developmental rate of engorged mosquito into egg laying  		& $2$			&	~\cite{CAILLY20127, EZANNO201539}					\\
$\sigma_A$      		& Sex-ratio at the emergence 								& $0.5$			&	~\cite{CAILLY20127, EZANNO201539}					\\
$\gamma_{em}$       & Mortality rate during adult emergence 						& $0.1$			&	~\cite{CAILLY20127, EZANNO201539}					\\
$\gamma_{A_{0}}$ 	& Egg laying rate 										& $f(T)$			&	~\cite{YU201828}					\\
$\zeta_0$ 			& Mortality rate of mosquito due to adulticide 					& $0.55$			&	~\cite{BHOWMICK2020110117}					\\
$\eta_L$                    & Mortality rate of larvae due to larvicide                                             & $0.6$	                 &	Assumed	                                                              \\
[1ex] 
 \hline
\end{tabular}
\caption{Parameters of the model \eqref{Eq1}, \eqref{Eq2} and \eqref{Eq3} and their corresponding values. 
In this context, $f(T)$ represents the function describing temperature dependence and the descriptions are included in the section \ref{Temperature_driven_parameters}.
$T$ describes temperature.
}
\label{table:2}
\end{table}

%%%%%%%%%%%%%%%%%%%%%%%%%%%%%%%%%%%%%%%%%%%%%%%%%%%%%%%%%%%%%%%%%%%%%%%%%%%%%%%%%%%%%%%%%%
%%%%%%%%%%%%%%%%%%%%%%%%%%%%%%%%%%%%%%%%%%%%%%%%%%%%%%%%%%%%%%%%%%%%%%%%%%%%%%%%%%%%%%%%%%
%%%%%%%%%%%%%%%%%%%%%%%%%%%%%%%%%%%%%%%%%%%%%%%%%%%%%%%%%%%%%%%%%%%%%%%%%%%%%%%%%%%%%%%%%%

\section{Temperature driven parameters}\label{Temperature_driven_parameters}

In this section, we have described the weather-driven model parameters, the external forcing functions driving the fecundity, the transition rate and mortality functions that govern the life-cycle dynamics of \textit{Culex} mosquitoes, adapting the model specifically to conditions in Cook County, Illinois, USA.
The current formulation includes \textit{Culex} dispersal and therefore it applies to a connected metapopulation model.
Given the considerable phenotypic variability within \textit{Culex} populations and the limited availability of parameter data, we have estimated parameter values based on expert knowledge of local \textit{Culex} biology and information drawn from the scientific literature ~\cite{CAILLY20127, EZANNO201539, YU201828, BHOWMICK2020110117, LAPERRIERE201199, Jia, ijerph10051698, BHOWMICK2020110117, Neil, Joac, 10.7554/eLife.58511}.
The forcing functions driving the parameters associated in the life-cycle of \textit{Culex} in our model depend on temperature ($T$) and precipitation ($P$), both varying with time.
We have utilised the daily mean values from 2000-2021 to run the simulations as illustrated in the section \ref{Modelsimulation}.
We have focused on \textit{Culex} biological processes strongly influenced by temperature-aquatic development, mortality, and maturation.
Our model reproduces seasonal population dynamics of aquatic and adult stages, with stage transitions governed by temperature fluctuations.
The development rate of stage $C$ is expressed as a temperature-dependent function $T(t)$ to capture seasonality.
Model parameters are detailed in ~\cite{BHOWMICK2024107346}, and graphical representations are provided in the Supplementary Information (SI).

%%%%%%%%%%%%%%%%%%%%%%%%%%%%%%%%%%%%%%%%%%%%%%%%%%%%%%%%%%%%%%%%%%%%%%%%%%%%%%%%%%%%%%%%%%
%%%%%%%%%%%%%%%%%%%%%%%%%%%%%%%%%%%%%%%%%%%%%%%%%%%%%%%%%%%%%%%%%%%%%%%%%%%%%%%%%%%%%%%%%%
%%%%%%%%%%%%%%%%%%%%%%%%%%%%%%%%%%%%%%%%%%%%%%%%%%%%%%%%%%%%%%%%%%%%%%%%%%%%%%%%%%%%%%%%%%
\section{Basic Offspring Number}\label{Basic_Offspring_Number}
The key parameter in population dynamics is the basic offspring number, $R_0$, defined as the average number of adult offspring produced by a single female \textit{Culex} during her lifetime \eqref{R0Single1}, \eqref{eq:R0_two_patch}.
This threshold value governs the stability of the model system \eqref{Eq1}, \eqref{Eq2} and \eqref{Eq3}.
If $R_0<1$, the system converges to the trivial, mosquito-free equilibrium (MFE), $T_0 = (E_0, L_0, P_0, A_0)=(0, 0, 0, 0)$, whereas $R_0 > 1$ yields a non-trivial equilibrium of the model system \eqref{Eq1}.
The quantity $R_0$ corresponds to the spectral radius of the next generation operator ~\cite{YANG_MACORIS_GALVANI_ANDRIGHETTI_WANDERLEY_2009, doi:10.1142/S0218339015500278}. 
To compute it, we linearise the model system \eqref{Eq1} and \eqref{Eq3} around the MFE and apply the Next Generation Matrix method ~\cite{doi:10.1098/rsif.2009.0386, VANDENDRIESSCHE200229}, obtaining explicit expressions for both single-patch and two-patch systems.
Further details regarding the NGM computations are provided in the SI.
Basic Offspring number without adulticide and larvicide:
\begin{equation} \label{R0Single1}
 R_{0_{1}} = 
\left(\frac{\alpha_A  \gamma_{A_{0}} }{ \beta_A }\right)
\left(\frac{\gamma_L \psi}  {\psi\gamma_L + \beta_E}\right)
 \left( \frac{\gamma_P} {\beta_L + \beta_W+ \gamma_P}\right)
  \left( \frac{ \sigma_A e^{- \gamma_{em}}\gamma_A} {\beta_P + \gamma_A}\right)
\end{equation}

Basic Offspring number with adulticide:
\begin{equation} \label{R0Single2}
R_{0_{1A}} = \left(\frac{\alpha_A  \gamma_{A_{0}} }{ \beta_A+\zeta }\right)
\left(\frac{\gamma_L \psi}  {\psi\gamma_L + \beta_E}\right)
 \left( \frac{\gamma_P} {\beta_L + \beta_W+ \gamma_P}\right)
  \left( \frac{ \sigma_A e^{- \gamma_{em}}\gamma_A} {\beta_P + \gamma_A}\right)
\end{equation}

Basic Offspring number with adulticide and larvicide:
\begin{equation} \label{R0Single3}
 R_{0_{1C}} =
 \left(\frac{\alpha_A  \gamma_{A_{0}} }{ \beta_A+\zeta }\right)
\left(\frac{\gamma_L \psi}  {\psi\gamma_L + \beta_E}\right)
 \left( \frac{\gamma_P} {\beta_L + \beta_W+ \gamma_P+\eta_L}\right)
  \left( \frac{ \sigma_A e^{- \gamma_{em}}\gamma_A} {\beta_P + \gamma_A}\right)
\end{equation}

\begin{align} 
R_0 &= 
\frac{1}{2(\beta_A + r_{12} + r_{21})}
\Big[
R_{0_{1}}(\beta_A + r_{21}) + R_{0_{2}}(\beta_A + r_{12}) \notag \\
&\quad +
\sqrt{
\big(R_{0_{1}}(\beta_A + r_{21}) - R_{0_{2}}(\beta_A + r_{12})\big)^2
+ 4\,R_{0_{1}}R_{0_{2}}\,r_{12}r_{21}
}
\Big]
\label{eq:R0_two_patch}
\end{align}

\noindent
where $R_{0_{i}}$ denotes the local basic offspring number in patch $i$, 
The expression in Eq.~\eqref{eq:R0_two_patch} captures the combined effect of local reproduction and 
migration on the metapopulation basic offspring number.

\paragraph{Biological interpretation}

Here, we explore the biological interpretations of the Basic Offspring Numbers ($R_0$ and $R_{0_{i}}$, when $i = 1, 2$) under different conditions.
When $r_{12} = r_{21} = 0$, there is no movement of mosquitoes between patches, and the metapopulation Basic Offspring Number simplifies to
\[
R_0 = \max\{ R_{0_{1}}, R_{0_{2}} \}
\]
As $r_{12}$ and $r_{21}$ increase, migration couples the dynamics of the two patches, producing an effective $R_0$ that reflects a weighted interaction between the local values.
In the limit of large symmetric migration ($r_{12} = r_{21} \to \infty$),
the two subpopulations behave as a single well-mixed population, and
\[
R_0 \to \frac{R_{0_{1}} + R_{0_{2}}}{2}
\]
Thus, dispersal can elevate the system-level reproductive potential when a high-risk patch is connected
to a low-risk one, emphasising the importance of coordinated control efforts across patches.
The biological meaning of the terms in \eqref{R0Single1} can be interpreted directly. 
During the non-diapausing phase, the expression for $R_{0_{i}}$, when $i = 1, 2$ can be expressed ecologically as the product of sequential survival fractions across life stages of \textit{Culex}: the proportion of eggs deposited that successfully hatch into larvae; the fraction of larvae that develop into pupae after exposure to water flushing and larvicide; the proportion of pupae maturing into adults in the terrestrial stage; and finally, the fraction of adults that survive natural mortality and adulticide.
%%%%%%%%%%%%%%%%%%%%%%%%%%%%%%%%%%%%%%%%%%%%%%%%%%%%%%%%%%%%%%%%%%%%%%%%%%%%%%%%%%%%%%%%%%%%%%%%%%%%%%%%%%%%%%%%%%%%%%%%%%%%%%%%%%%%%%%%%%%%%%%%%%%%%%%%%%%%%%%%%%%%%%%%%%%%%%%%%%%%%%%%%%%%%%%%%%%%%%%%%%%%%%%%%%%%%%%%%%%%%%%%%%%%%%%%%%%%%%%%%%%%%%%%%%%%%%%%%%%%%%%%%%%%%%%%%%%%
\section{Model simulation}\label{Modelsimulation}

We have modelled \textit{Culex} abundance across life stages ($E$, $L$, $P$, $A$), focusing on daily averages of adult ($A$) for epidemiological comparison with field collections ~\cite{BHOWMICK2025103163,10.1093/jme/tjae041, 10.1371/journal.pone.0332621}. 
We have numerically solved the metapopulation model systems \eqref{Eq2} and \eqref{Eq3} for a two-patch configuration, implemented in R ~\cite{R} for demonstration purposes, using PRISM weather data (2000–2021) ~\cite{PRISMClimateGroup} as environmental forcing inputs after following the work outlined in ~\cite{BHOWMICK2025103163, BHOWMICK2024107346, BHOWMICK2020110117, LAPERRIERE201199}.

We have run the simulations that spanned ten years to reach equilibrium and  solved using an explicit Euler scheme with a daily step as we have used the daily PRISM weather data. 
Initial conditions are $10^6$ eggs and $10^2$  adults on January $1^{\text{st}}$  ~\cite{CAILLY20127}, with overwintering mosquitoes maintained at $1\%$ of the initial value to avoid extinction ~\cite{BHOWMICK2020110117, LAPERRIERE201199}. 
For results in Sections~\ref{Model_validation} and \ref{Evaluating_spraying_strategies}, we have used 2018 weather data (Figure~\ref{fig:WeatherData}); further details are provided in the SI.
For demonstration purposes, we have considered a two-patch modelling framework to carry out our simulations. 
The insights gained from this framework can be qualitatively extended to more complex, multi-patch systems.
Adult dispersal between patches is modelled as bidirectional movement with rates $r_{12}$ and $r_{21}$, creating a coupled two-patch metapopulation modelling framework ~\cite{Neil}. 
Control interventions are represented as time-dependent increases in mortality for specific life stages: larvicide application increases larval mortality ($\eta_L(t)$) with a fixed mortality , while ULV spraying introduces an additional adult mortality term ($\zeta(t)$). 
The resulting system of weather-driven ODE system allows us of exploring the transient and steady-state responses to different interventions under realistic thermal and dispersal conditions.
Further details regarding the initial conditions are provided in the SI.

%%%%%%%%%%%%%%%%%%%%%%%%%%%%%%%%%%%%%%%%%%%%%%%%%%%%%%%%%%%%%%%%%%%%%%%%%%%%%%%%%%%%%%%%%%%%%%%%%%%%%%%%%%%%%%%%%%%%%%%%%%%%%%%%%%%%%%%%%%%%%%%%%%%%%%%%%%%%%%%%%%%%%%%%%%%%%%%%%%%%%%%%%%%%%%%%%%%%%%%%%%%%%%%%%%%%%%%%%%%%%%%%%%%%%%%%%%%%%%%%%%%%%%%%%%%%%%%%%%%%%%%%%%%%%%%%%%%%

\section{Sensitivity analysis}\label{Sensitivity_analysis}
\begin{figure}[H]
\centering
\subfloat[Single patch]{\includegraphics[width=0.5\textwidth]{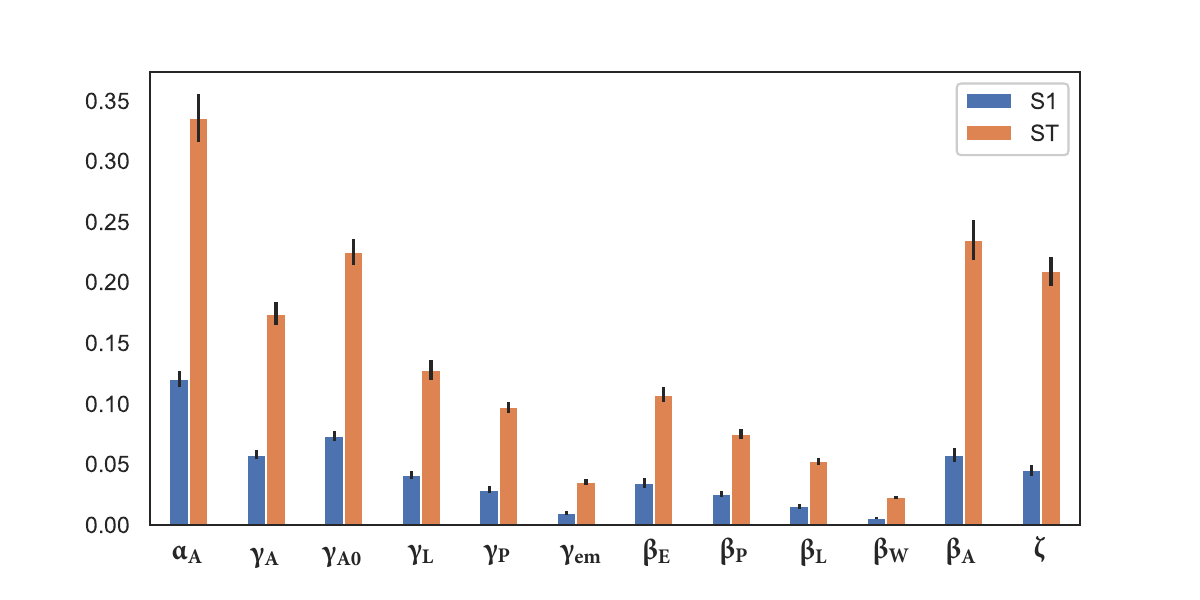 }\label{fig:s1}}
\hfill
\subfloat[Two patch]{\includegraphics[width=0.45\textwidth]{  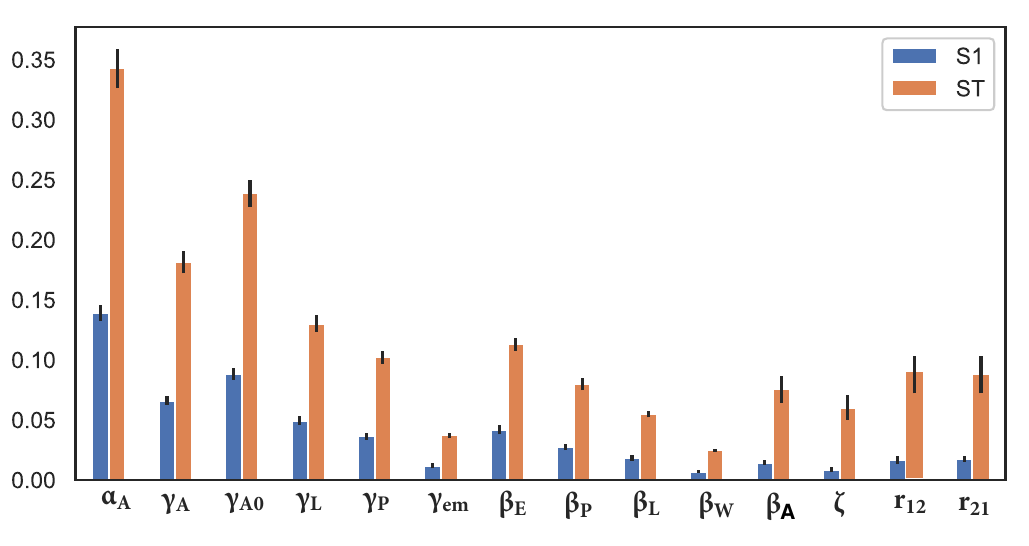 }\label{fig:s2}}
\caption{Sobol indices.
}\label{fig:Sens1Inte}
\end{figure}

\begin{figure}[H]
\centering
\subfloat[Single patch]{\includegraphics[width=0.45\textwidth]{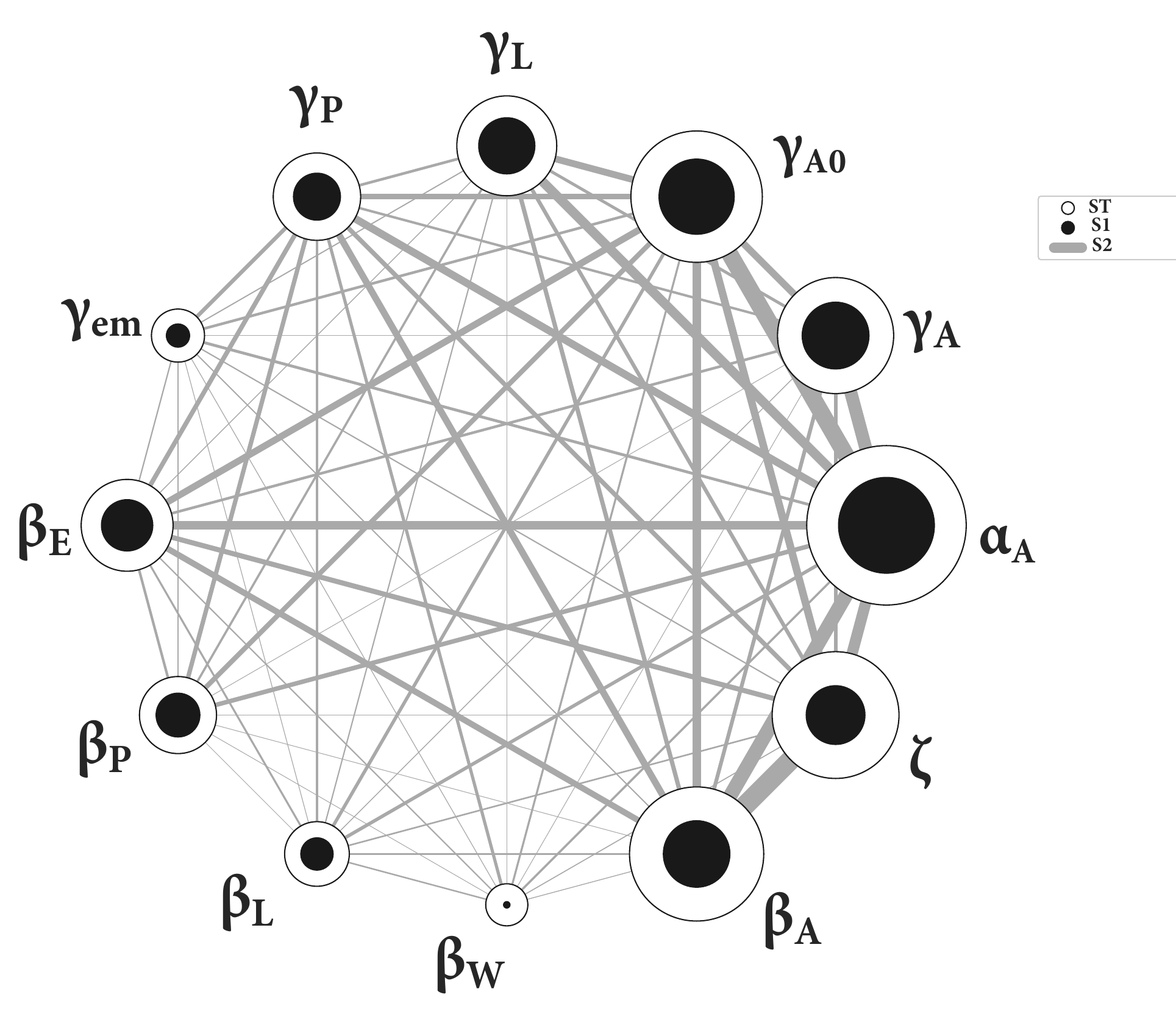}\label{fig:s11}}
\hfill
\subfloat[Two patch]{\includegraphics[width=0.45\textwidth]{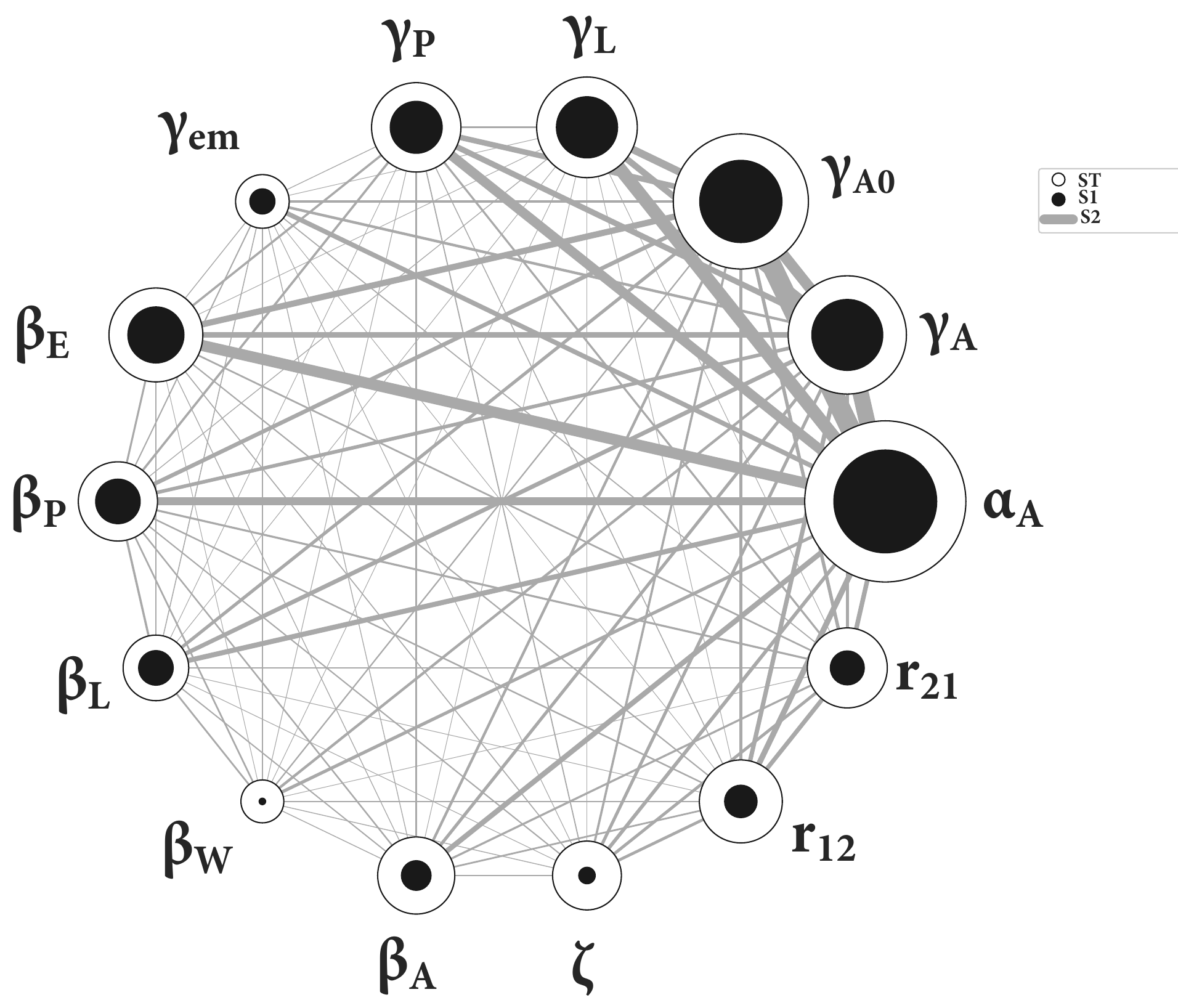}\label{fig:s21}}
\caption{Sobol indices of second-order interactions.
}\label{fig:Sens2Inte}
\end{figure}

We have performed a global sensitivity analysis (GSA) to evaluate the influence of model parameters on the Basic Offspring numbers ($R_{0}$ and $R_{0_{1}}$) for both the single-patch and two-patch models \eqref{Eq1} and \eqref{Eq2}. 
Using SALib ~\cite{Iwanaga2022, Herman2017}, we have computed first-order (S1), second-order (S2), and total (ST) Sobol indices (Figure~\ref{fig:Sens1Inte}, \ref{fig:Sens2Inte}). 
The S1 index describes the proportion of output variance attributable to each parameter individually, thus enabling a direct comparison of parameter importance between the two model structures.
The second-order indices (S2) quantifies the effects arising from the interaction between two parameters, such as $x_i$ and $x_j$ and in our case, these parameters correspond to the Basic Offspring number for the single and two-patch models, $R_{0}$ and $R_{0_{1}}$, respectively.
The total effects (ST) describe both first-order effects and all higher-order interactions for a given parameter. 
Sensitivity indices quantify each input’s fractional contribution to output variance and reveal parameter interactions, here evaluated for the mean Basic Offspring numbers of the single- and two-patch models.
By default, $95\%$ confidence intervals were reported for each index.
The GSA has included defining simulation size, selecting parameters, setting ranges, and assigning uniform and Poisson distributions.
Means and variances are computed per ~\cite{Iwanaga2022, Herman2017}.
First-order effects are calculated by fixing each parameter in turn, and total effects by including both individual and interaction contributions.
We first select and visualise the total (ST) and first-order (S1) indices for each input in both the single-patch \eqref{Eq1}  and two-patch models \eqref{Eq2}.
The S1 index characterises the individual contribution of each input to the output variance, while the S2 values quantify the second-order interactions among inputs. 
To better illustrate these interactions, we have employed a network-based visualisation method.
Here, the sizes of the ST and S1 circles depict the normalised importance of the variables in relation to the response functions ($R_{0_{i}}$ and $R_{0}$, when $i = 1, 2$).

Figure~\ref{fig:s1} shows that $\alpha_A$, $\gamma_{A_{0}}$, $\beta_A$, $\zeta$ and $\gamma_A$ have the highest S1 and ST indices, together accounting for over $90\%$ of the variance in $R_{0_{1}}$, when considering both individual effects and interactions with other parameters. 
These are followed in importance by $\gamma_L$, $\beta_E$, $\gamma_P$, $\beta_P$, $\gamma_{em}$ and $\beta_W$, respectively.
Figure~\ref{fig:s2} shows that $\alpha_A$, $\gamma_{A_{0}}$, $\gamma_A$ have the highest S1 and ST indices, together explaining over $65\%$ of the variance in $R_0$ when accounting for both individual effects and interactions with other parameters. 
These are followed in importance by $\gamma_L$, $\gamma_P$, $r_{12}$, $r_{21}$, $\beta_E$, $\beta_P$, $\beta_A$, $\zeta$, $\gamma_{em}$ and $\beta_W$, respectively.
According to the Sobol sensitivity index, S$i$, where $i=1,2$, 
$\alpha_A$, $\beta_A$, $\zeta$, $\gamma_{A_{0}}$, $\gamma_{A}$,  $\gamma_{L}$ and $\beta_E$ are identified as the primary contributors to the variance of the Basic offspring number ($R_{0_{i}}$, when $i = 1, 2$).

According to the figure~\ref{fig:s11}, $\zeta$ exhibits strong interactions with $\beta_A$, $\alpha_A$, $\gamma_{A_{0}}$, followed by $\beta_E$, $\gamma_L$, $\gamma_P$ and $\gamma_{A}$.
Based on the Sobol sensitivity indices, $\alpha_A$, $\gamma_{A_{0}}$, $\gamma_{A}$,  $\gamma_{L}$, $\beta_E$, $r_{12}$ and $r_{21}$ are identified as the primary contributors to the variance of the Basic Offspring Number ($R_{0}$).
Figure~\ref{fig:s21} shows that $\zeta$ exhibits neither strong first-order nor second-order interactions with other mosquito parameters, yet it has a relatively moderate magnitude of total-order interaction whereas $\alpha_A$ has a strong interaction with $\gamma_{A_{0}}$, $\gamma_{A}$, $\gamma_L$, $\gamma_P$ and $\beta_E$. 
In the two-patch model, $r_{12}$ and $r_{21}$  exhibit relatively strong interactions with $\alpha_A$, $\gamma_{A_{0}}$ and $\gamma_{A}$, as well as mutual interactions with each other, and display larger total-order interactions compared to $\zeta$.

Figure~\ref{fig:s1} indicates that $R_{0_{i}}$ is most sensitive to the number of eggs laid by \textit{Culex}, and the natural mortality rate of adult \textit{Culex},, followed by the egg-laying rate, mortality rate due to adulticide (ULV spray), developmental rate from pupae to adults, developmental rate from larvae to pupae, and the egg mortality rate, respectively.
Similarly, the Figure~\ref{fig:s2} illustrates the influence of different mosquito parameters on $R_{0}$. 
Here, $R_{0}$ is also sensitive to the number of eggs laid by \textit{Culex}, the egg-laying rate, developmental rates from pupae to adults and from larvae to pupae, as well as the mortality rates of eggs, adult \textit{Culex}, and pupae. However, unlike in the case of $R_{0_{1}}$, the influence of the mortality rate due to adulticide (ULV spray) ($\zeta$) is not as apparent.
We also notice that $R_{0}$ is sensitive to the dispersal rates between two neighbouring patches ($r_{12}$ and $r_{21}$). 
Figures~\ref{fig:Sens1Inte} and \ref{fig:Sens2Inte} convey several important insights relevant to the \textit{Culex}  lifecycle and mosquito abatement planning, particularly in the context of mosquito dispersal.
The parameter distributions used to calculate the Sobol indices index is presented in the SI.

%%%%%%%%%%%%%%%%%%%%%%%%%%%%%%%%%%%%%%%%%%%%%%%%%%%%%%%%%%%%%%%%%%%%%%%%%
%%%%%%%%%%%%%%%%%%%%%%%%%%%%%%%%%%%%%%%%%%%%%%%%%%%%%%%%%%%%%%%%%%%%%%%%%
%%%%%%%%%%%%%%%%%%%%%%%%%%%%%%%%%%%%%%%%%%%%%%%%%%%%%%%%%%%%%%%%%%%%%%%%%

\section{Model validation}\label{Model_validation}
We have fitted a weather-driven, ODE-based mechanistic model \eqref{Eq2},  \eqref{Eq3} (two-patch) against NWMAD trap data of 2018 (Figure~\ref{fig:TrapData} (d)), evaluating its ability to reproduce observed \textit{Culex} abundance by comparing simulations with time-series trap counts after following the authors in ~\cite{CHEN2023106837}.
We have employed the sequential MCMC method to fit the model-predicted counts of female adult \textit{Culex} mosquitoes to the observed counts recorded in each trap.
On each trap day, we assumed that the observed mosquito count followed a Poisson distribution, with a mean proportional to the mosquito population predicted by the model for the entire community. 
Let $C_i$ denote the trap count on day $i$; then
\begin{equation}\label{MCMC1}
C_i \approx \text{Poisson}(A_i,\psi)
\end{equation}
where $\psi$ represents the capture rate of adult \textit{Culex} mosquitoes.
The expected trap count is proportional to the proportion of blood-seeking female adults and the trap’s efficiency in capturing them.
The sequential MCMC method samples from the posterior distributions of the model parameters by evaluating the likelihood function, thereby estimating parameter values consistent with the observed data by following the likelihood function
\begin{equation}\label{MCMC2}
\prod_{i \, \text{for all trap days}}\frac{[A_i\psi]^{C_i}e^{-A_i,\psi}}{C_i !}.
\end{equation}

\begin{figure}[H]
\centering
\subfloat[Trap site: Maryville Academy (MV)]{\includegraphics[width=0.45\textwidth]{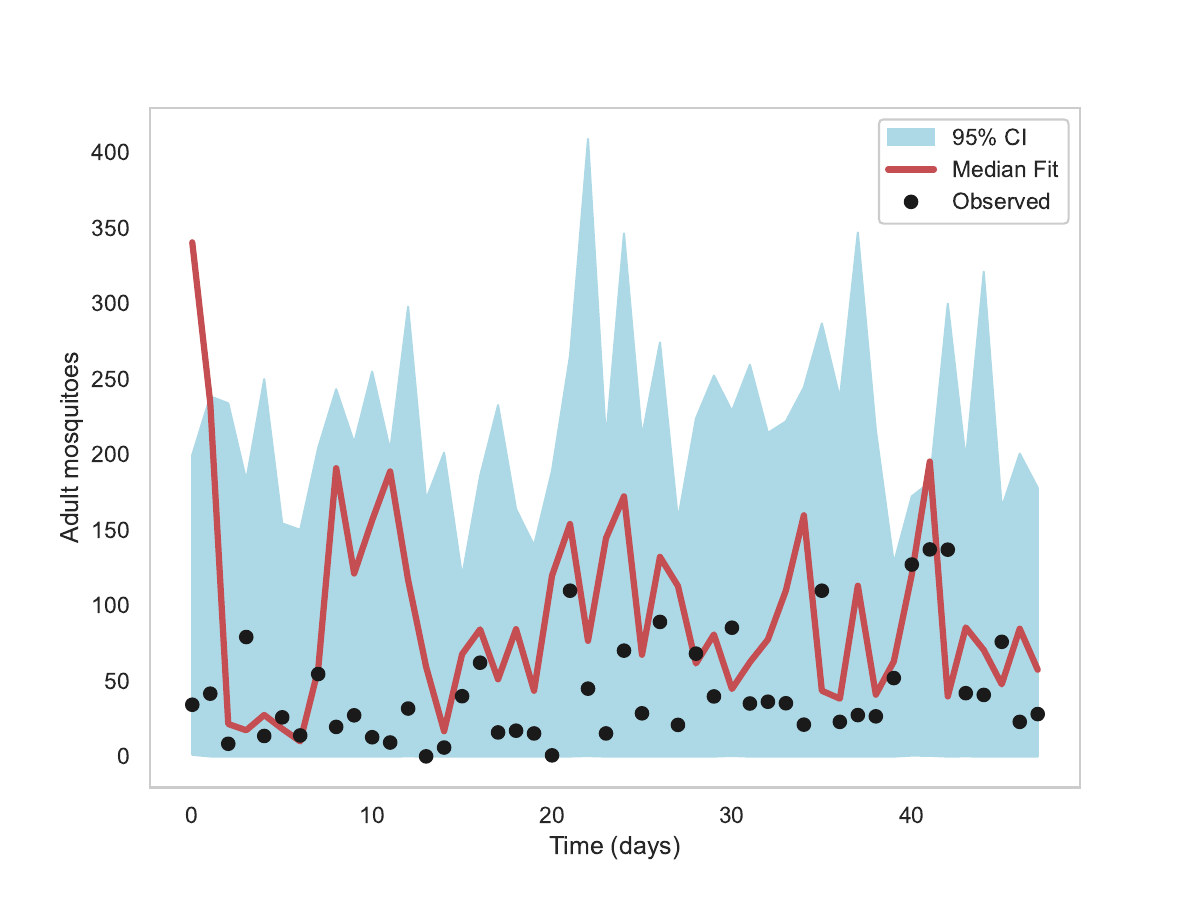}\label{fig:fit1}}
\hfill
\subfloat[Trap site:  All Saints Cemetery (AS) ]{\includegraphics[width=0.45\textwidth]{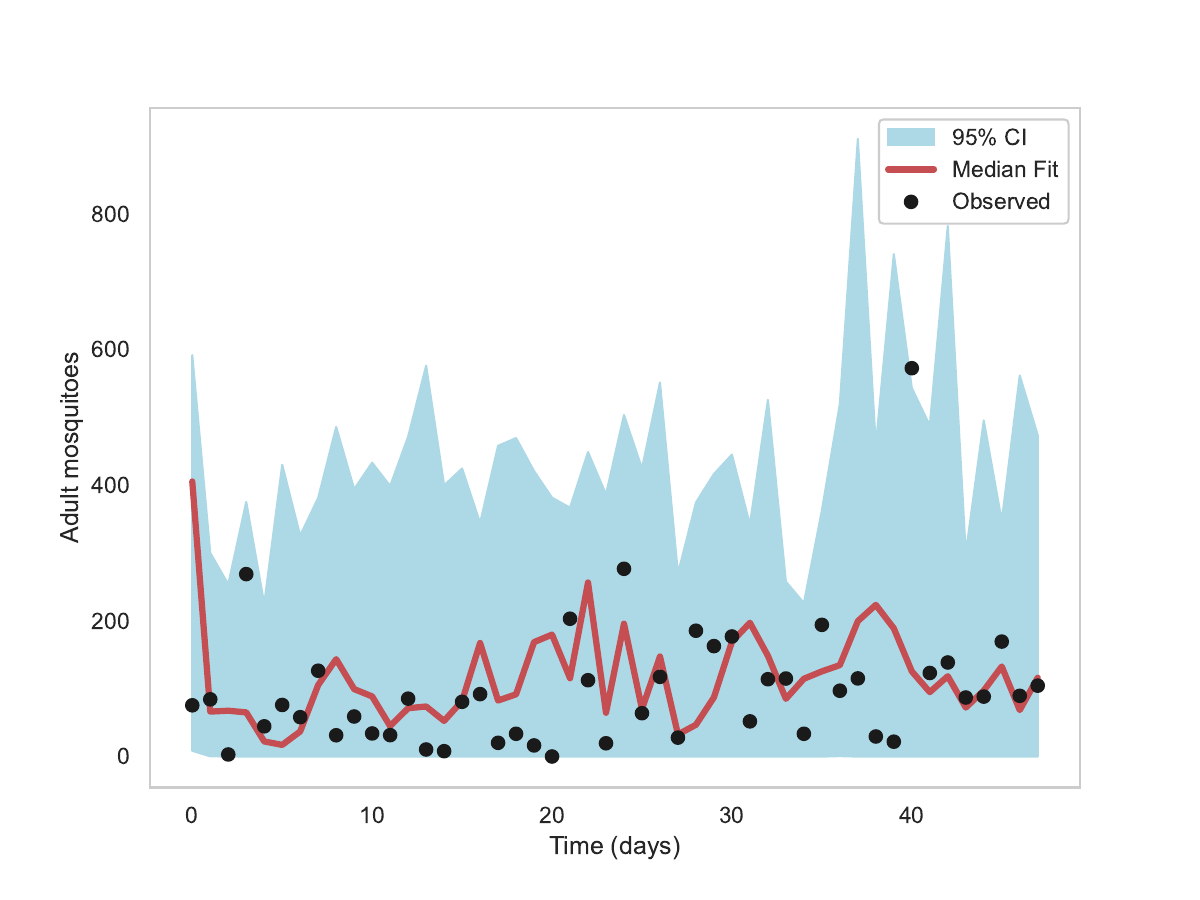}\label{fig:fit2}}
\caption{
A comparison between the model simulation generated from our model as described in \eqref{Eq2} and trap data. 
Here, we showed the calibrated simulations with
two adjacent  trap data sets for the demonstration purpose. 
The black dots were the collected trap data (\ref{fig:TrapData} (c) and (d)) and the red lines were the calibrated solution from
the model system \eqref{Eq2} for a two-patch system.
The colored bands indicate that the trap count of \textit{Culex} obtained on each trap day bounded within the band with a $95\%$ probability.
}\label{fig:Modelfit}
\end{figure}
The details about the simulation process is described in the \ref{Modelsimulation}.
We have utilised the dataset from $2018$ from two neighbouring sites: AS = All Saints Cemetery and MV = Maryville Academy (Figure~\ref{fig:TrapData} (d)).
MV, the treatment site, had 5 weeks of adulticide treatments. AS, the control site, had 1 adulticide treatment that are included during the model fit.
Please note that for this data set, there were times when the traps were left on all weekend, so those counts seem very high because they are a multi-day collection event. 
The spray event on 3/7/2018 - both control and treatment sites were being sprayed. 
All other spray events were just the treatment site (MV).
%%%%%%%%%%%%%%%%%%%%%%%%%%%%%%%%%%%%%%%%%%%%%%%%%%%%%%%%%%%%%%%%%%%%%%%%%%%%%%%%%%%%%%%%%%
%%%%%%%%%%%%%%%%%%%%%%%%%%%%%%%%%%%%%%%%%%%%%%%%%%%%%%%%%%%%%%%%%%%%%%%%%%%%%%%%%%%%%%%%%%
%%%%%%%%%%%%%%%%%%%%%%%%%%%%%%%%%%%%%%%%%%%%%%%%%%%%%%%%%%%%%%%%%%%%%%%%%%%%%%%%%%%%%%%%%%

\section{Evaluation of spraying strategies}\label{Evaluating_spraying_strategies}

Mosquito abatement depends on chemical, biological, and environmental strategies to suppress mosquito populations, with adulticide spraying remaining a widely used method for rapidly reducing adult mosquitoes and lowering disease risk ~\cite{10.1093/jme/tjad088, 10.1093/jme/tjae041, EZANNO201539}.
Across Cook County,Illinois, USA, integrated programs that combine larvicides with ULV spraying are commonly implemented to strengthen overall control efforts ~\cite{10.1371/journal.pone.0332621, 10.1093/jme/tjad088, BHOWMICK2025103163}.
Formulating the functional representation of $\zeta$ (the rate of adulticide effectiveness contributing to mosquito mortality) was of particular importance, as described in Section \ref{Multipatch_mosquito_dynamics_with_larvicide_and_adulticide}.
Since ULV applications were conducted exclusively during the summer months ~\cite{10.1093/jme/tjad088, EZANNO201539}, we have utilised a step function approach, following the methodology outlined in ~\cite{BHOWMICK2024107346, BHOWMICK2025103163}. 
The step function $\zeta(t)$ characterised the mosquito mortality rate resulting from ULV treatment, representing a proportional reduction in the population. 
This choice was motivated by the fact that ULV spraying delivers an immediate, short-term impact rather than a gradual decline, making a step function a parsimonious yet realistic representation of its effect. 
Using this formulation, we evaluated the effectiveness of different adulticide spraying strategies within a single season and examined how different strategies have influenced the relative abundance and the Basic Offspring number of \textit{Culex} mosquitoes in a two-patch model.

Several strategies and techniques are employed to apply adulticides effectively ~\cite{10.2987/19-6848.1, BHOWMICK2025103163, EZANNO201539, Demers}. 
Mathematical modeling and simulations can play a crucial role in planning and optimising mosquito abatement strategies, including the application of adulticides-larvicide in multi-patch settings. 
The NWMAD implements a variety of control approaches, and we have simulated selected strategies based on expert knowledge to facilitate a comparative analysis ~\cite{10.2987/19-6848.1, 10.1093/jme/tjz083}. 
Specifically, we have evaluated several distinct spraying strategies and compared them to a baseline scenario without adulticide application, using both single-patch and two-patch models. 
These strategies were denoted as $S_i$, with $S_0$ representing the absence of adulticide treatment. In our simulations, we have assumed a daily ULV treatment effectiveness of $55\%$, which was incorporated as an additional mortality factor affecting the adult mosquito population. 
Furthermore, we also have conducted simulations combining larvicide and adulticide applications to evaluate the effectiveness of integrated control strategies with a daily larvicide treatment effectiveness of $60\% (\eta_L)$.

%%%%%%%%%%%%%%%%%%%%%%%%%%%%%%%%%%%%%%%%%%%%%%%%%%%%%%%%%%%%%%%%%%%%%%%%%%%%%%%%%%%%%%%%%%
%%%%%%%%%%%%%%%%%%%%%%%%%%%%%%%%%%%%%%%%%%%%%%%%%%%%%%%%%%%%%%%%%%%%%%%%%%%%%%%%%%%%%%%%%%
%%%%%%%%%%%%%%%%%%%%%%%%%%%%%%%%%%%%%%%%%%%%%%%%%%%%%%%%%%%%%%%%%%%%%%%%%%%%%%%%%%%%%%%%%%

\begin{table}[H]
\centering
\begin{tabular}{||c c c c ||} 
 \hline
 Spray start time & Interval  & Number of times & Strategy name \\ [0.5ex] 
 \hline\hline
No spray & 0 & 0 & $S_0$   \\ 
Mid July &  Once a week & $3$ weeks & $S_1$\\
Late July &  Once a week & $5$ weeks & $S_2$\\
First week of July &  Twice a week & $3$ weeks & $S_3$\\ 
First week of June &  Once a week & $5$ weeks & $S_4$\\
Mid May&  Once a week & $5$ weeks & $S_5$\\
First week of June &  Everyday & $3$ consecutive days & $S_6$\\
Third week of August &  Everyday & $3$ consecutive days & $S_7$\\
[1ex] 
 \hline
\end{tabular}
\caption{
NWMAD prescribed various spray regimes employed in our model \eqref{Eq1},  \eqref{Eq1Modified} to assess optimal strategies.
}
\label{table:A}
\end{table}

%%%%%%%%%%%%%%%%%%%%%%%%%%%%%%%%%%%%%%%%%%%%%%%%%%%%%%%%%%%%%%%%%%%%%%%%%%%%%%%%%%%%%%%%%%
%%%%%%%%%%%%%%%%%%%%%%%%%%%%%%%%%%%%%%%%%%%%%%%%%%%%%%%%%%%%%%%%%%%%%%%%%%%%%%%%%%%%%%%%%%
%%%%%%%%%%%%%%%%%%%%%%%%%%%%%%%%%%%%%%%%%%%%%%%%%%%%%%%%%%%%%%%%%%%%%%%%%%%%%%%%%%%%%%%%%%

\subsection{Difference between single-patch and two-patch model}\label{Difference_between_single_patch_and_two_patch_model}
\begin{figure}[H]
\centering
\subfloat[]{\includegraphics[width=0.5\textwidth, height=5.5cm]{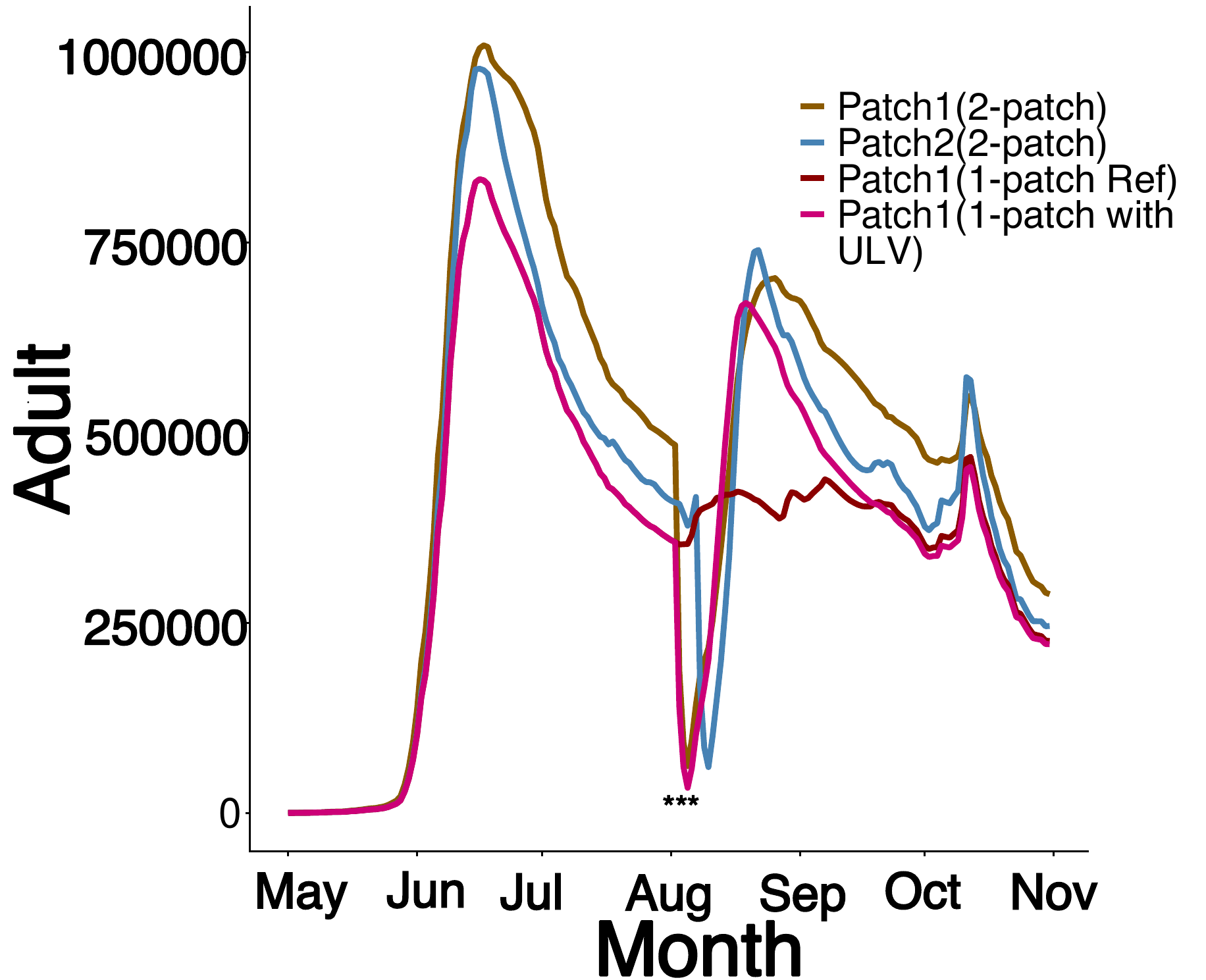}\label{fig:2PatchAbundance}}
\hfill
\subfloat[]{\includegraphics[width=0.5\textwidth, height=5.5cm]{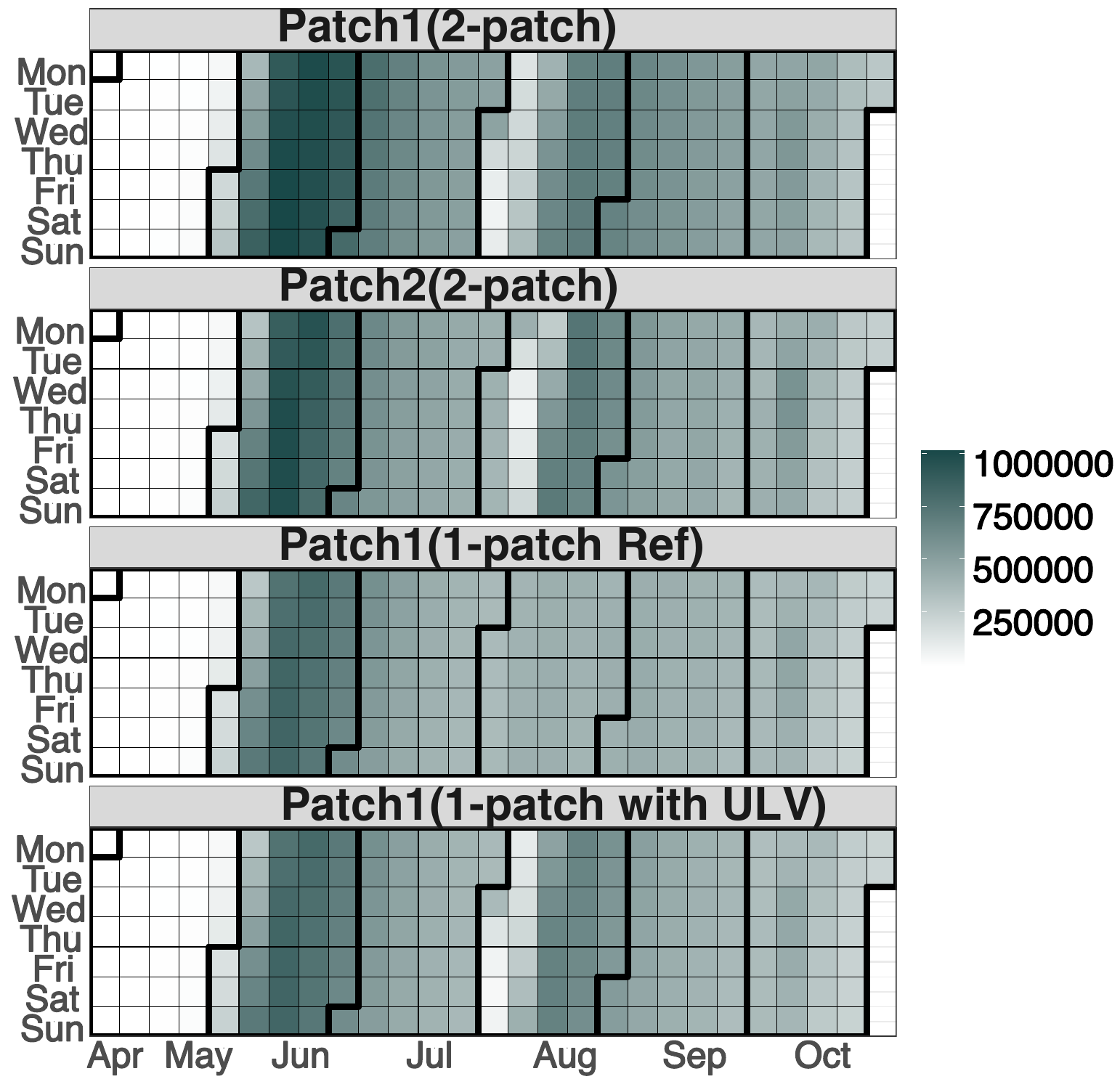}\label{fig:2PatchCalender}}
\caption{
We have displayed the relative abundance of \textit{Culex} mosquitoes observed throughout the simulation period under the $S_7$ control strategy as described in the Table \ref{table:A} across four different simulation scenarios: (i) a single patch without ULV spraying, (ii) a single patch with ULV spraying, (iii) Patch 1, and (iv) Patch 2 in a two-patch metapopulation model in the Figure \ref{fig:2PatchAbundance}.
Figure \ref{fig:2PatchCalender} has displayed  the relative abundance of \textit{Culex} under four different simulation scenarios: (i) a single patch without ULV spraying, (ii) a single patch with ULV spraying, (iii) Patch 1, and (iv) Patch 2 in a two-patch metapopulation model and they have been cumulatively simulated on a daily basis
and displayed in a calendar plot, which also marks the days when adulticide has been applied, as indicated in Table \ref{table:A} and shown in the Figure \ref{fig:2PatchAbundance}.
}\label{fig:2PatchAbundanceComparison}
\end{figure}

In this section, we describe the difference between the effects of different control strategies employed by the NWMAD on a single-patch system and two-patch model system connected by the mosquito dispersal.
We have employed the abatement strategy $S_7$ for the demonstration purpose as described in the Table \ref{table:A}.
One can readily notice that in a single-patch system, ULV spray acts solely as an additional mortality term that directly suppresses the local adult \textit{Culex} population.
The outcome dynamics are relatively straightforward as the relative abundance of adult \textit{Culex} decreases sharply following the spray events and afterwards the population bounces back rapidly as new adults emerge from pupae, possibly as depicted in the Figure \ref{fig:2PatchAbundanceComparison}.
Therefore, the overall nonzero equilibrium point i.e. the relative abundance of \textit{Culex} depends on the balance between recruitment, maturation in the presence of ULV spray, larvicide in a weather-driven setting and the dynamics are radically different for a single-patch and a two-patch model.
If the ULV spray is applied continuously in the summer season, then ultimately the system converges towards a steady state or demonstrates damped oscillations determined by temperature-driven recruitment and maturation.
We can postulate that as there is no immigration, population recovery entirely depends on local reproduction, thus making control effects appear more effective with the condition of temperature-driven sensitive parameters.

\subsubsection{Assessing the effect of abatement strategies on relative abundance of \textit{Culex}}

To test our hypothesis, we have described and visually represented the relative abundance of \textit{Culex} mosquitoes under four different scenarios: (i) a single patch without ULV spraying, (ii) a single patch with ULV spraying, (iii) Patch 1, and (iv) Patch 2 in a two-patch metapopulation model.
To perform the simulation, we have employed $S_7$ ULV spraying scheme as outlined in the Table \ref{table:A}.

A weather-driven two-patch model inducts spatial feedback through the dispersal ($(r_{12}$, $r_{21}$)) of adult \textit{Culex} mosquitoes. 
When ULV spray is applied in one habitat patch of \textit{Culex} reduces local abundance but can potentially trigger a compensatory dynamic due to dispersal from the untreated or not equally treated neighbouring habitat patch. 
This re-invasion effect might be able to delay the suppression or even maintain persistent low-level populations despite having an intensive control on a local level.

\subsubsection{Assessing the effect of  abatement strategies on Basic Offspring Numbers}
\begin{figure}[H]
\centering
\subfloat[]{\includegraphics[width=0.5\textwidth, height=5.5cm]{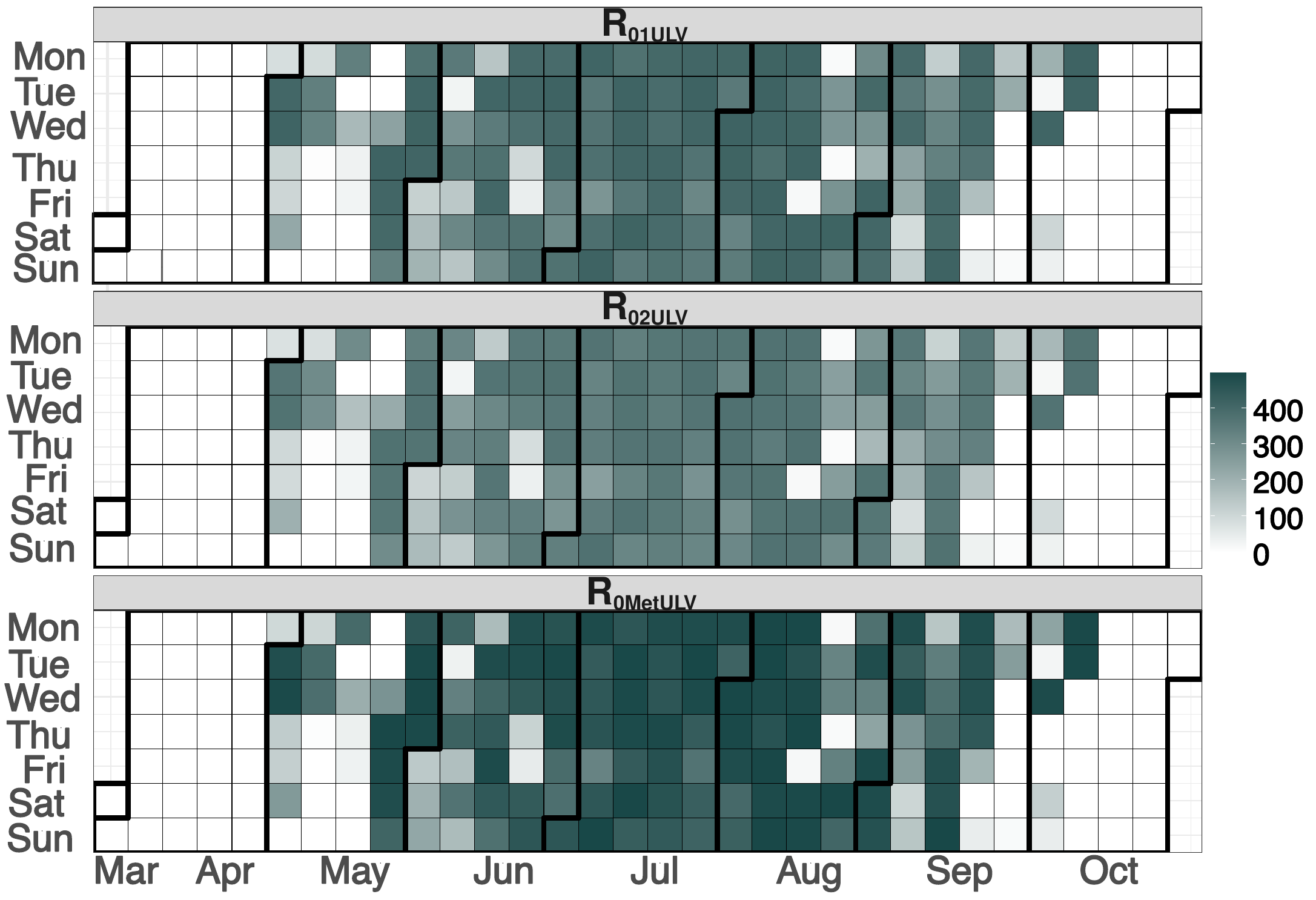}\label{fig:R0Heatmap1}}
\hfill
\subfloat[]{\includegraphics[width=0.5\textwidth, height=5.5cm]{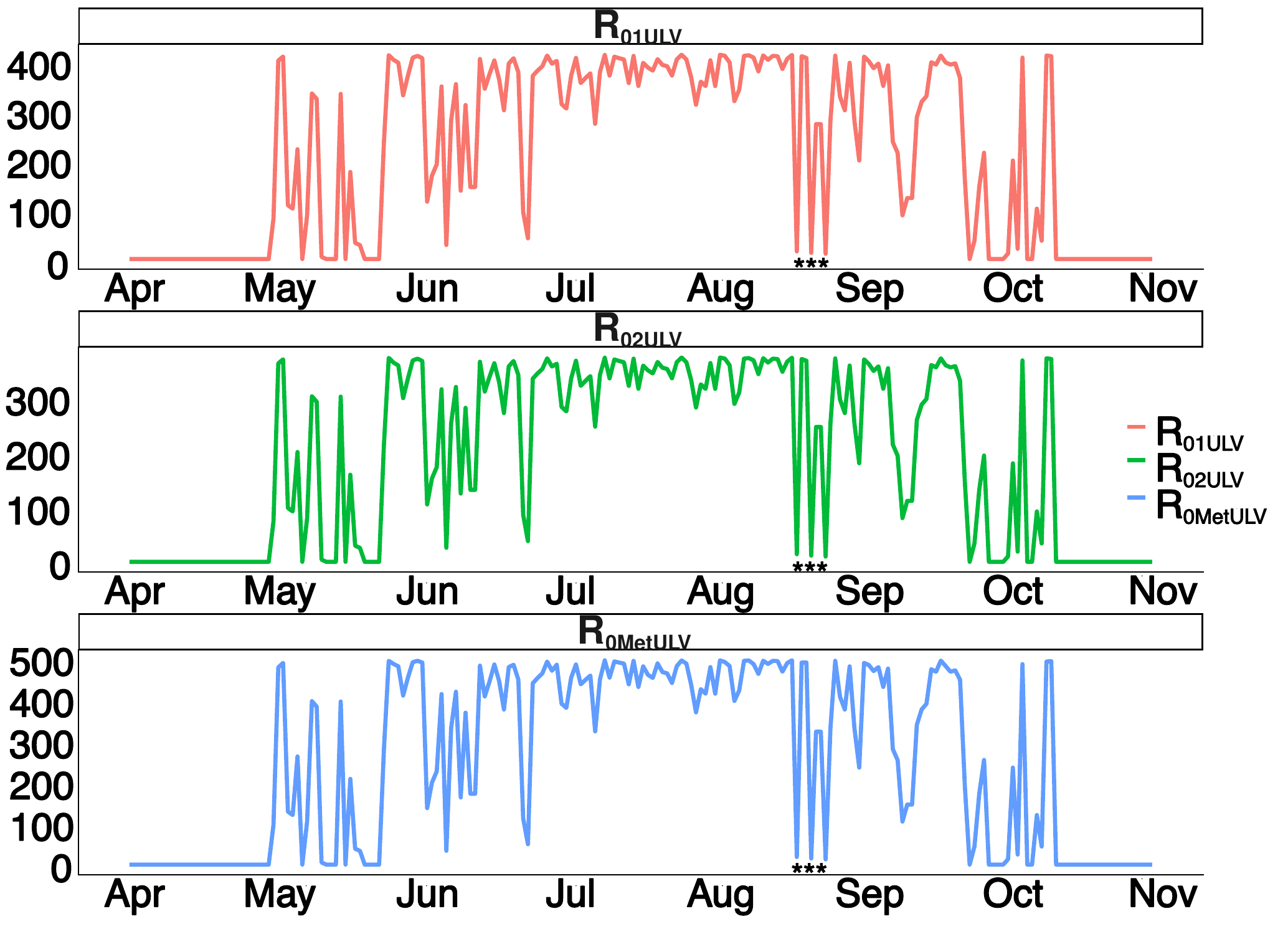}\label{fig:R0Heatmap2}}
\caption{
We have depicted a comparison of the magnitudes of the Basic Offspring numbers for Patch 1 ($R_{0_{1}}$) and Patch 2 ($R_{0_{2}}$), as well as the overall $R_0$ of the metapopulation connected through \textit{Culex} dispersal rates ($r_{12}$, $r_{21}$) in the Figure \ref{fig:R0Heatmap2} under the control scheme $S_7$ as described in the Table  \ref{table:A}.
The figure \ref{fig:R0Heatmap1}  has illustrated the Basic Offspring Numbers ($R_{0_{1}}$, $R_{0_{2}}$,  $R_0$), under the $S_7$ ULV spray scheme. 
The values of $R_1$, $R_2$, and $R_0$ have been cumulatively simulated on a daily basis and displayed in a calendar plot, which also marks the days when adulticide has been applied, as indicated in Table \ref{table:A} and shown in the Figure  \ref{fig:R0Heatmap2}.
}\label{fig:R0Comparison}
\end{figure}

In this section, we assess how ULV adulticide spraying influences the Basic Offspring numbers ($R_{0_{1}}$, $R_{0_{2}}$, and $R_0$) in the two-patch mosquito population model. 
We have implemented the spraying strategy $S_7$ to examine its effect on the Basic Offspring numbers of the individual patches ($R_{0_{1}}$) and ($R_{0_{2}}$) as well as the metapopulation-level ($R_0$).
Figure~\ref{fig:R0Comparison} illustrates the impact of ULV spraying on the basic offspring numbers of each patch ($R_{0_{1}}$) and ($R_{0_{2}}$) and on the overall two-patch metapopulation ($R_0$) model, which are interconnected through adult \textit{Culex} ($r_{12}, r_{21}$) dispersal. When ULV spray is applied, the magnitudes of $R_{0_{1}}$ and $R_{0_{2}}$ temporarily decline due to the increased adult mortality; however, they recover rapidly as the untreated aquatic stages continue to replenish the adult \textit{Culex} population.

The reduction in the two-patch metapopulation model ($R_0$) is less pronounced than that observed in the single-patch scenario (Figure~\ref{fig:R0Comparison}). 
Our simulation suggests that the efficacy of ULV spraying alone may often be overestimated when it is considered as the sole mosquito abatement strategy, as migration and larval recruitment can potentially facilitate rapid population recovery across connected patches.
To address this limitation, the following section we explore the combined application of ULV spray and larvicide, and this  simultaneously targets both the adult and aquatic stages, thereby offering a more sustained reduction in \textit{Culex} mosquito abundance and a stronger suppression of across the two-patch system.

\subsection{Assessing the effect of different ULV spray strategies on the relative abundance of \textit{Culex} in a two-patch model}
In this section, we have simulated the effect of ULV spray  on the relative abundance of \emph{Culex} mosquitoes within the weather-driven two-patch model framework.
We have implemented the spraying strategies employed by the NWMAD as described in the Table \ref{table:A} to examine its effect on the relative abundance of \textit{Culex} population in a two-patch model model.
\begin{figure}[H]
\centering
\subfloat[Patch 1]{\includegraphics[width=0.5\textwidth, height=5.5cm]{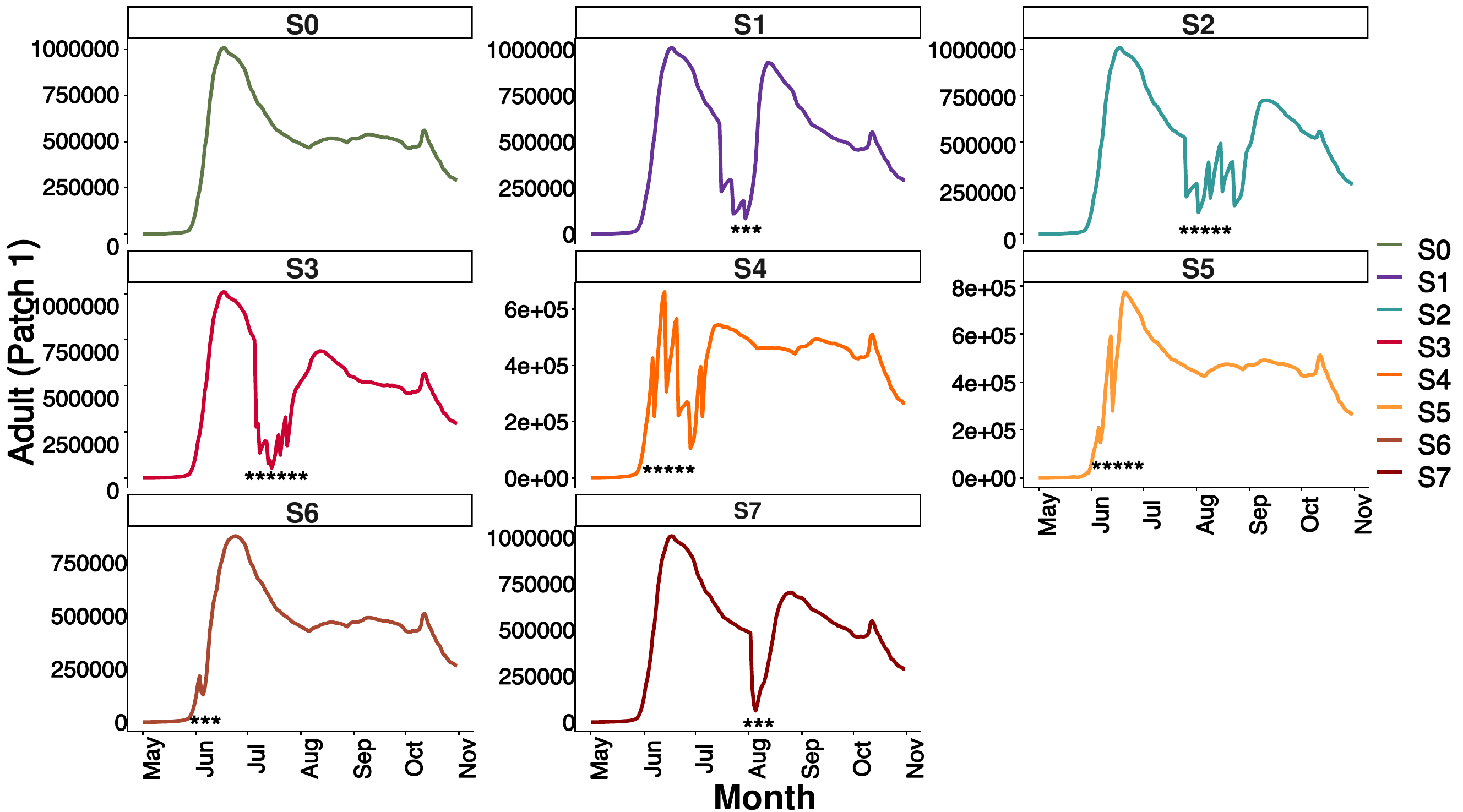}\label{fig:Comparisontwopatch1}}
\hfill
\subfloat[Patch 2]{\includegraphics[width=0.5\textwidth, height=5.5cm]{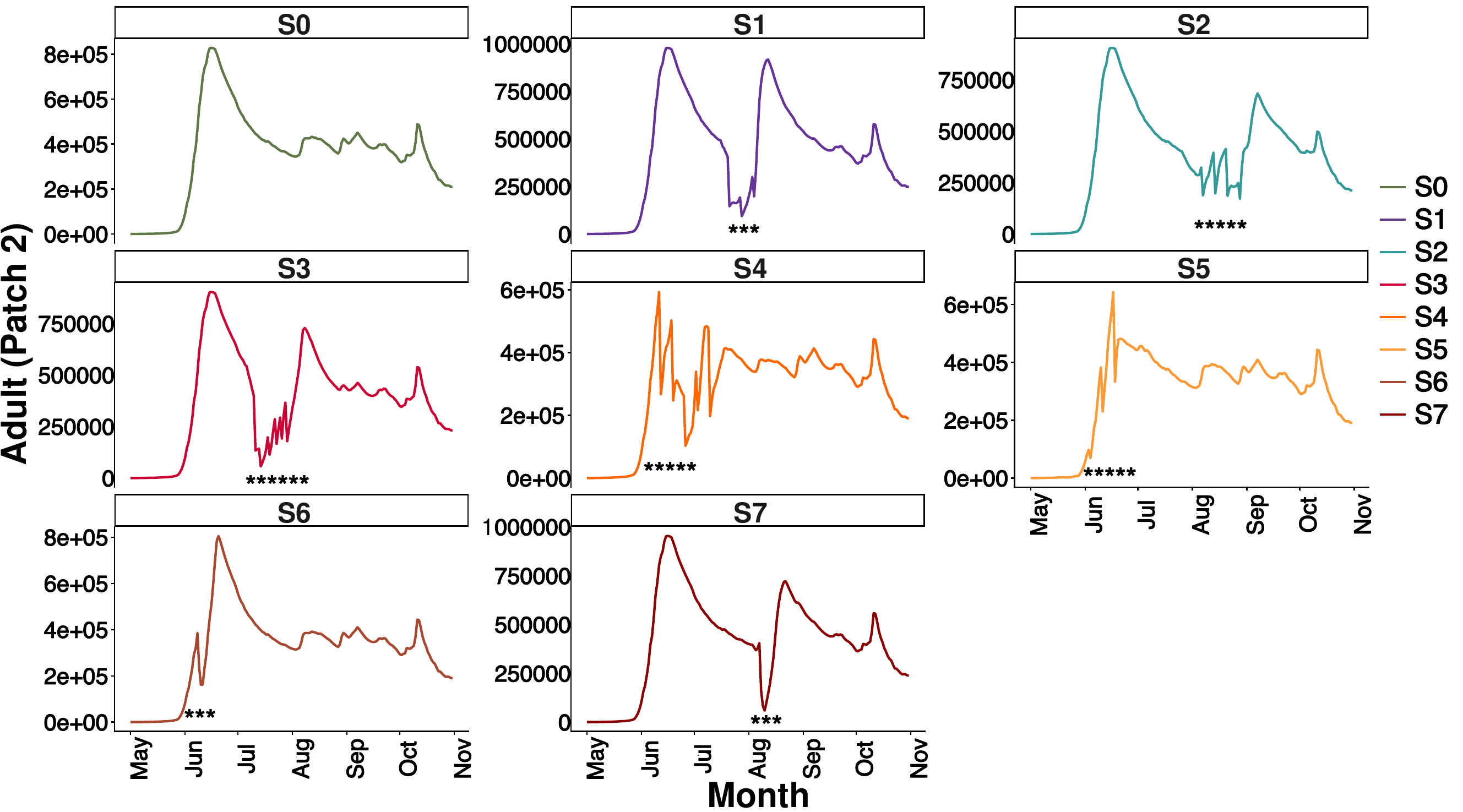}\label{fig:Comparisontwopatch2}}
\caption{
We have shown the relative abundance of \textit{Culex} observed throughout the simulation period under different control strategies of two patches.
Figures \ref{fig:Comparisontwopatch1} and \ref{fig:Comparisontwopatch2} represent the relative abundance of \textit{Culex} of Patch 1 and Patch 2 respectively.
The stars have indicated the days when adulticide has been applied, following the schedule outlined in the Table  \ref{table:A}.
We have also displayed the relative abundance of \textit{Culex} mosquitoes in two different patches connected through dispersal in the Figures \ref{fig:Comparisontwopatch1} and 
\ref{fig:Comparisontwopatch2}.
}\label{fig:two-patchCulexabundanceComparison}
\end{figure}

\begin{figure}[H]
\centering
\subfloat[Patch 1]{\includegraphics[width=0.5\textwidth, height=5.5cm]{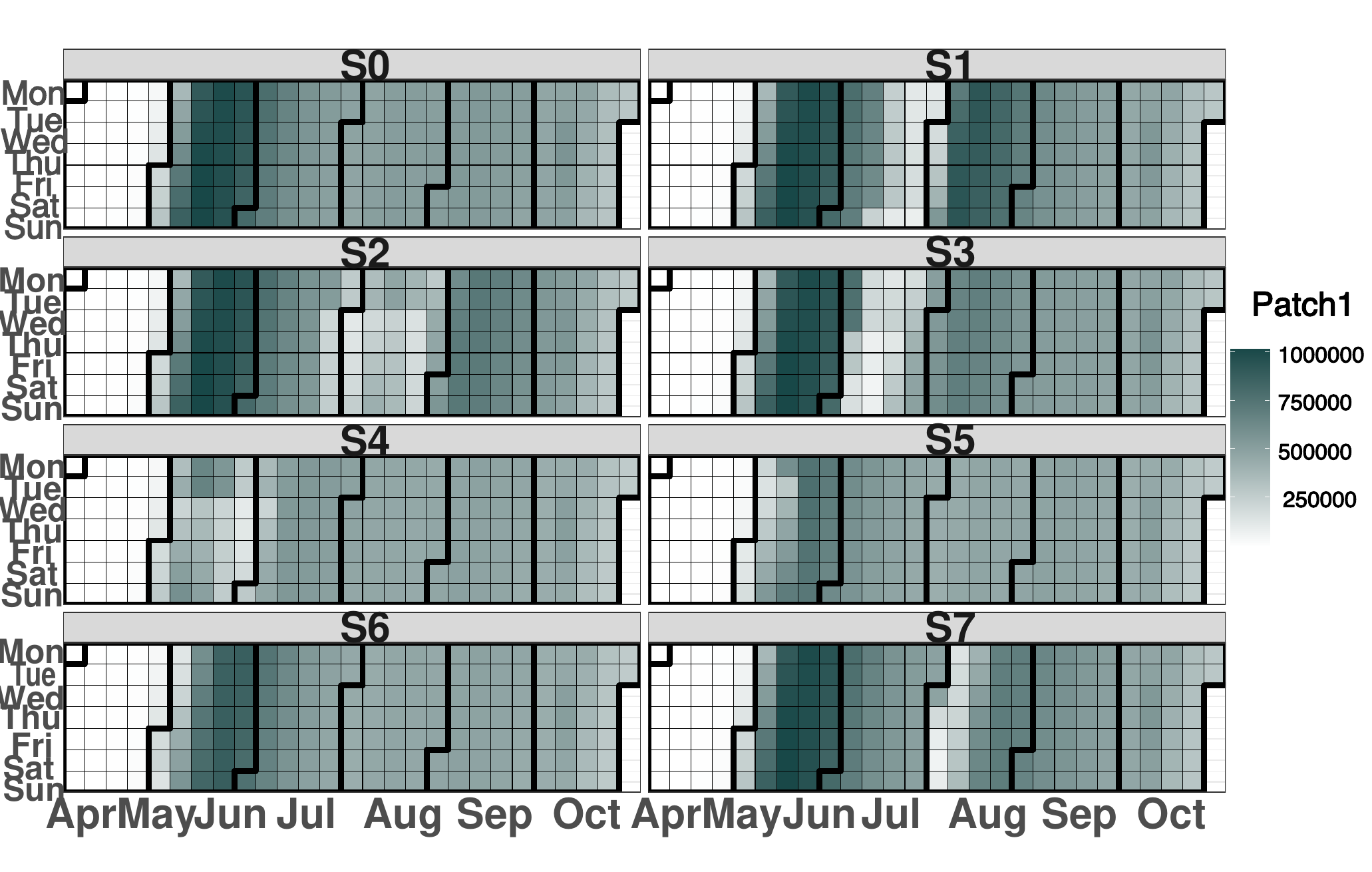}\label{fig:Comparisontwopatch1Calendar}}
\hfill
\subfloat[Patch 2]{\includegraphics[width=0.5\textwidth, height=5.5cm]{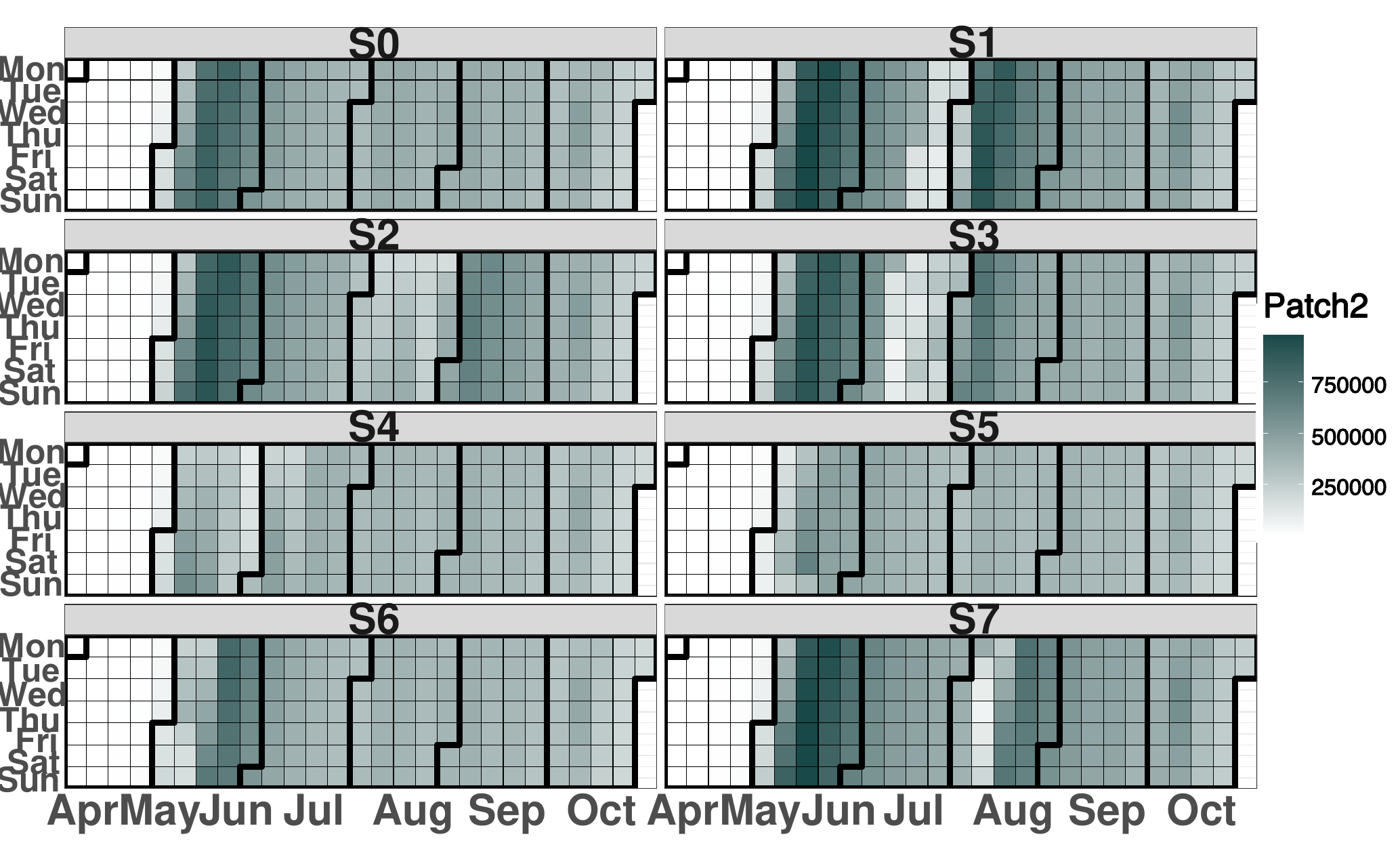}\label{fig:Comparisontwopatch2Calendar}}
\caption{
Daily \textit{Culex} abundance has been cumulatively simulated and presented in a calendar plot, which also includes the days when adulticide has been applied, as listed in Table  \ref{table:A} and shown in the Figure \ref{fig:two-patchCulexabundanceComparison}.
Figures \ref{fig:Comparisontwopatch1Calendar} and \ref{fig:Comparisontwopatch2Calendar} represent the simulated cumulative abundance of \textit{Culex} of Patch 1 and Patch 2 respectively.
}\label{fig:two-patchCulexabundanceComparisonCalendar}
\end{figure}

It is evident from Figures \ref{fig:two-patchCulexabundanceComparison} and \ref{fig:two-patchCulexabundanceComparisonCalendar} that ULV spraying conducted early in the season produces more favorable outcomes compared to applications performed during mid or late summer.

\subsection{Assessing the combined effect of ULV spray and larvicide on the relative abundance of \textit{Culex} in a two-patch model}
In this section, we have simulated the combined impact of ULV spray and larvicide on the relative abundance of \emph{Culex} mosquitoes and the basic offspring number ($R_0$) within the weather-driven two-patch model framework.
In the weather-driven two-patch mosquito population model \eqref{Eq3}, both larvicide and ULV spray interventions are included as extra mortality terms acting on distinct life stages. 
The larvicide primarily targets the aquatic immature stages (eggs, larvae, pupae) while increasing the mortality rate ($\eta_L$) of larval stage and the ULV spray increases adult mortality instantaneously adulticide efficacy term, $\zeta$.

\begin{figure}[H]
\centering
\subfloat[]{\includegraphics[width=0.5\textwidth, height=5.cm]{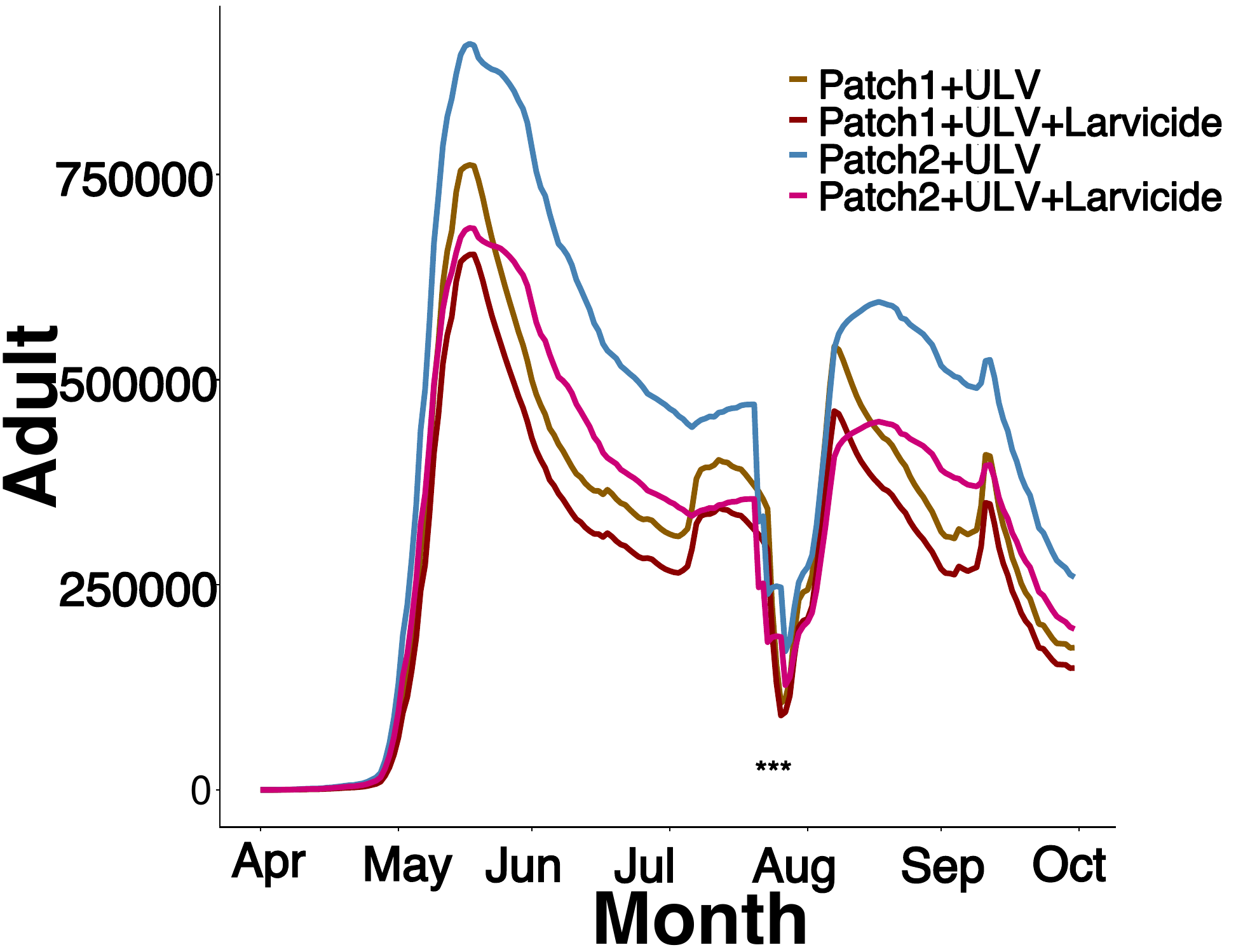}\label{fig:ULV_Larvicide1}}
\hfill
\subfloat[]{\includegraphics[width=0.5\textwidth, height=5.cm]{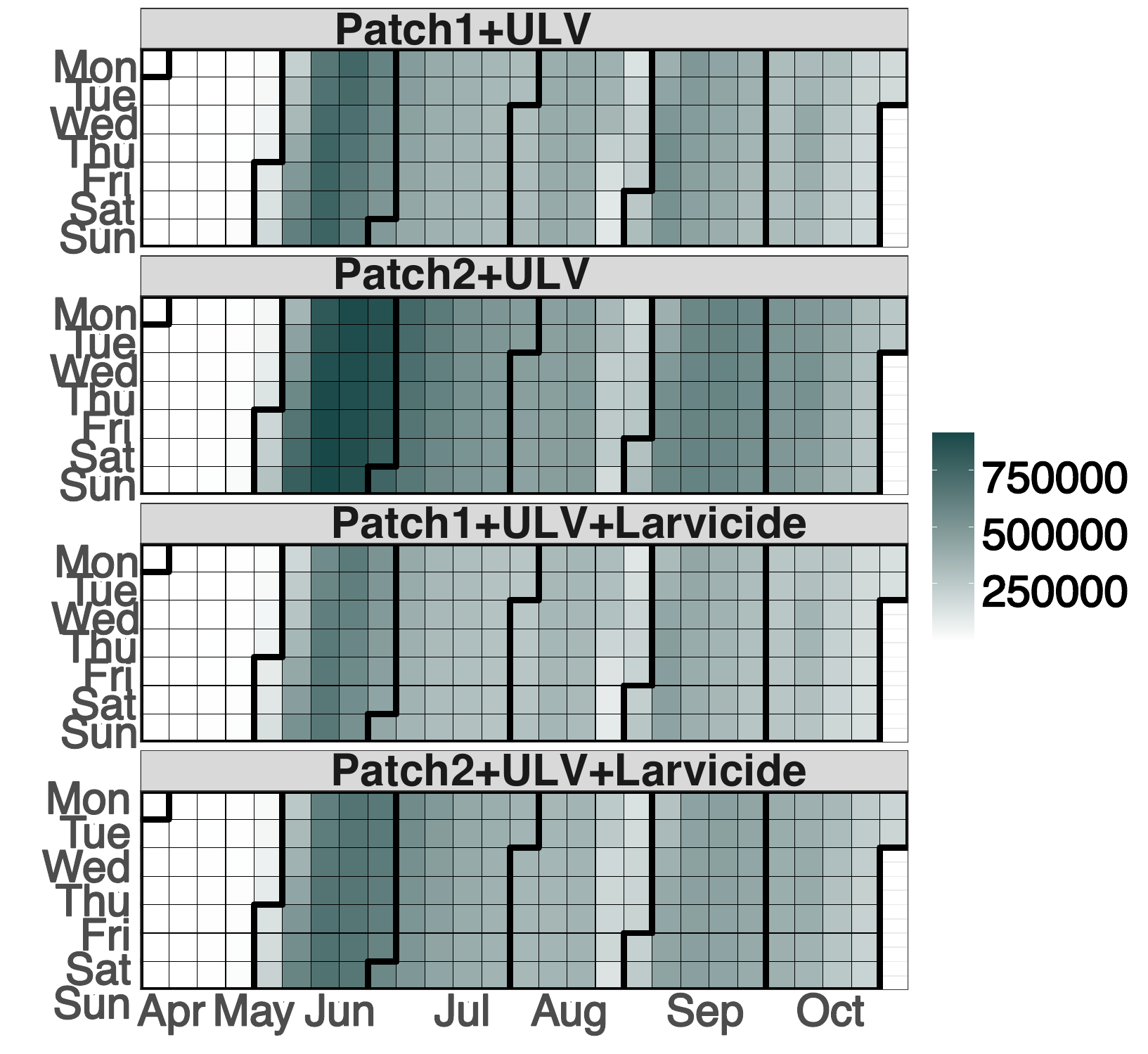}\label{fig:ULV_Larvicide2}}
\caption{
We have displayed the relative abundance of mosquitoes observed throughout the simulation period under the $S_7$ control strategies. 
The stars in the Figure \ref{fig:ULV_Larvicide1},  have indicated the days when adulticide was sprayed, as per the schedule outlined in the Table \ref{table:A}.
In the Figure \ref{fig:ULV_Larvicide2} daily mosquito abundance was cumulatively simulated and presented in a calendar plot, which includes the days when adulticide was applied, as listed in the Table  \ref{table:A} and shown in the Figure \ref{fig:ULV_Larvicide1}.
}\label{fig:ULVLarvicideTimeSeries}
\end{figure}

\begin{figure}[H]
\centering
\subfloat[]{\includegraphics[width=0.5\textwidth, height=5.5cm]{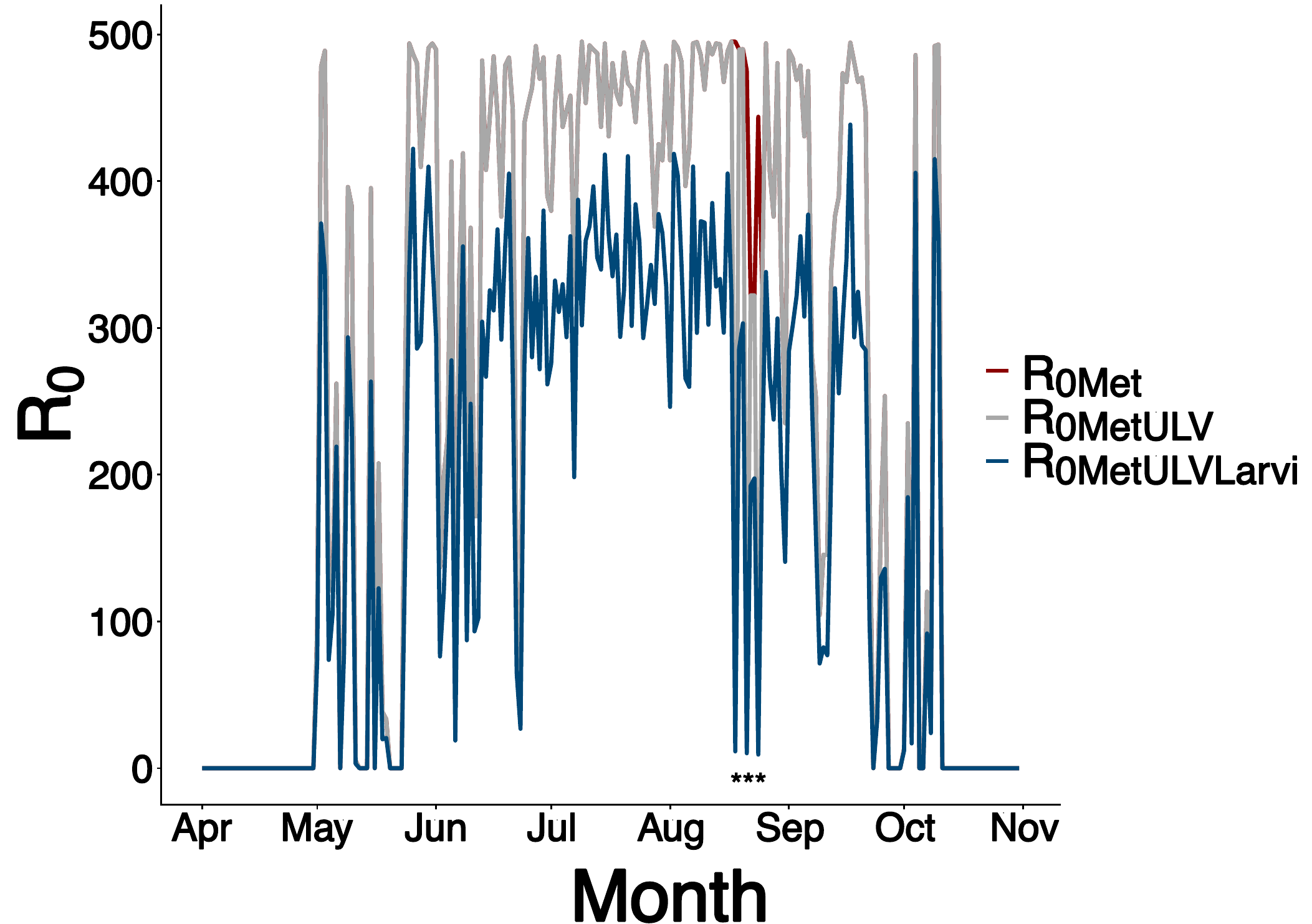}\label{fig:ULV_LarvicideR01}}
\hfill
\subfloat[]{\includegraphics[width=0.5\textwidth, height=6cm]{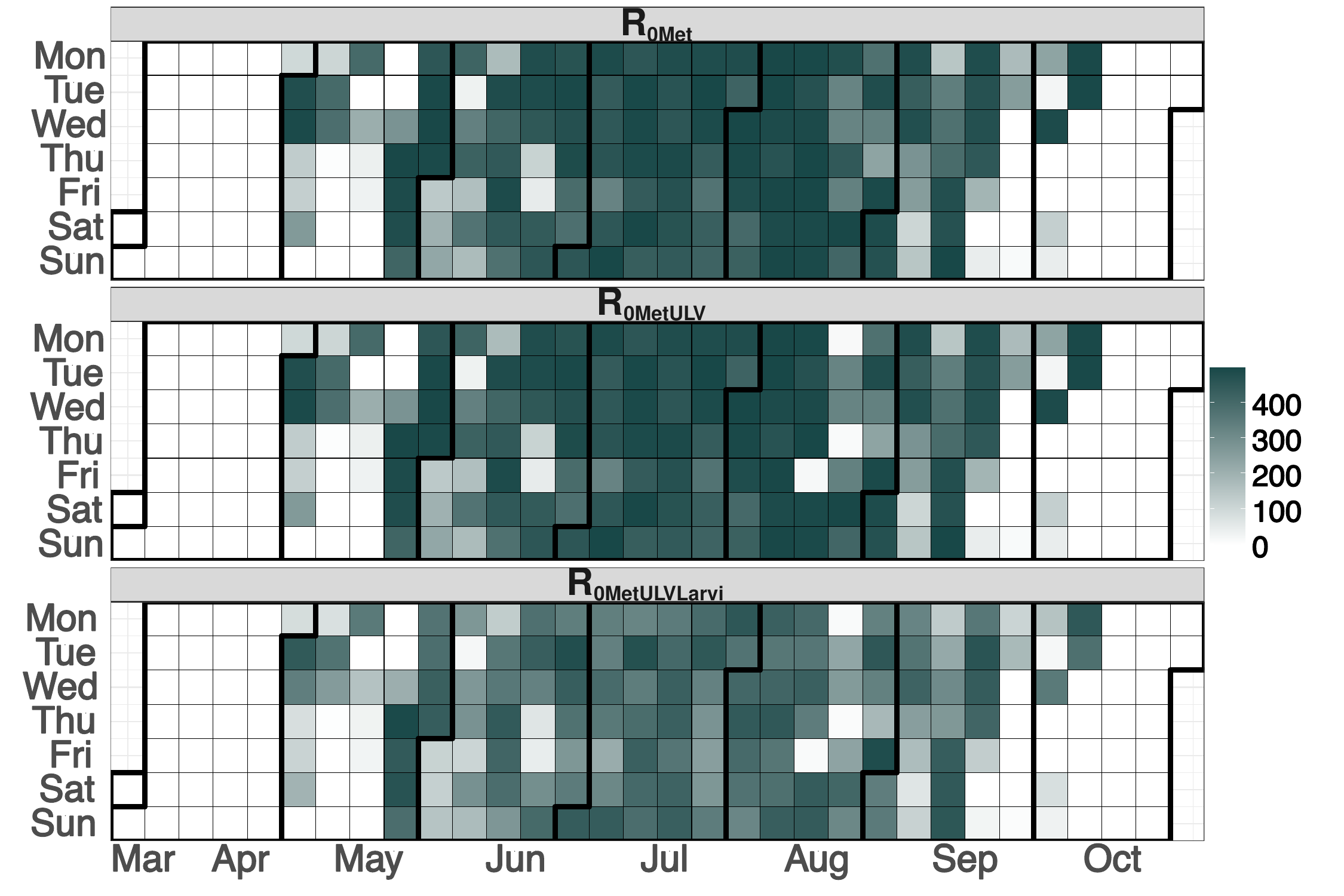}\label{fig:ULV_LarvicideR02}}
\caption{
We have presented the Basic Offspring Number $R_0$ observed over the simulation period for without any control strategy, under the $S_7$ ULV spray regime and in a combination of $S_7$ ULV spray regime and larvicide. 
The stars in the Figure \ref{fig:ULV_LarvicideR01} have marked the days when adulticide was applied, following the schedule outlined in the Table  \ref{table:A}.
The figure \ref{fig:ULV_LarvicideR02} has illustrated the Basic Offspring number, $R_0$, without any control strategy, under the $S_7$ ULV spray regime and in a combination of $S_7$ ULV spray regime and larvicide. 
The daily cumulative values of $R_0$ have been simulated and visualised in a calendar plot, which has also marked the days when adulticide has been applied according to the $S_7$ ULV spray schedule, as outlined in Table  \ref{table:A} and illustrated in Figure \ref{fig:ULV_LarvicideR01}.
}\label{fig:ULVLarvicideR0}
\end{figure}

We have illustrated in Figures \ref{fig:ULVLarvicideTimeSeries} and \ref{fig:ULVLarvicideR0} the differences between scenarios where we have implemented only the $S_7$ strategy and those where the $S_7$ spray is combined with larvicide in a two-patch metapopulation.
Figure \ref{fig:ULVLarvicideTimeSeries} presents the time series of relative \textit{Culex} abundance, showing that the synchronous application of ULV spray and larvicide markedly reduces the adult mosquito population.
Similarly, Figure \ref{fig:ULVLarvicideR0} demonstrates that the combined interventions lower the overall magnitude of the Basic Offspring Number ($R_0$) of two-patch metapopulation, thus highlighting the enhanced effectiveness of an integrated control.

During the simulation, when both control interventions are applied synchronously or in coordination, the system \eqref{Eq3} experiences a double pressure on \textit{Culex} population renewal: larval recruitment into the adult compartment is suppressed, and existing adult \textit{Culex} mosquitoes are removed at relatively higher rates. This has resulted in a marked reduction in the overall adult \textit{Culex}abundance $A_i$ ($i = 1, 2$) and consequently lowers the potential for recolonisation through migration between these two patches. Mathematically, the simulation has shown us that this could lead us to a reduction in the Basic Offspring Number ($R_{0_{1}}, R_{0_{1}}$) for the two-patch metapopulation model compared to a scenario with a single control measure.

%%%%%%%%%%%%%%%%%%%%%%%%%%%%%%%%%%%%%%%%%%%%%%%%%%%%%%%%%%%%%%%%%%%%%%%%%%%%%%%%%%%%%%%%%%
%%%%%%%%%%%%%%%%%%%%%%%%%%%%%%%%%%%%%%%%%%%%%%%%%%%%%%%%%%%%%%%%%%%%%%%%%%%%%%%%%%%%%%%%%%
%%%%%%%%%%%%%%%%%%%%%%%%%%%%%%%%%%%%%%%%%%%%%%%%%%%%%%%%%%%%%%%%%%%%%%%%%%%%%%%%%%%%%%%%%%

\section{Discussion and Conclusion}
Our weather-driven mechanistic mathematical model of \textit{Culex} population dynamics in a patchy environment provides new insights into how spatial heterogeneity-mosquito dispersal and vector control interventions interact to shape mosquito abundance. 
By explicitly incorporating patch connectivity and adult mosquito migration, the simplified mosquito lifecycle model has tried to capture a relatively realistic ecological dynamics compared to single-patch frameworks. 
This is particularly significant in urban and semi-urban landscapes, where breeding habitats are spatially fragmented but connected through adult \textit{Culex} dispersal. 
This weather-driven patchy model encompasses the entire simplified mosquito life cycle and incorporates the diapause mechanism in a straightforward manner. 
Our model has demonstrated the first mechanistic approach to understanding the population dynamics of \textit{Culex} mosquitoes in a patchy environment and providing a closed-form expression for the Basic Offspring number ($R_0$) and its relationship with temperature and different interventions.
Our simulations align closely with data on the number of mosquito pools trapped of two-neighbouring patches, confirming the consistency and accuracy of our weather-driven ODE model as described in the section \ref{Model_validation}.\par

Our simulation-based study informs evidence-based strategies for public health authorities and policymakers combating vector-borne diseases in the presence of mosquito dispersal. 
In addition, our current patchy model puts an effort to bridge the gap between theoretical modelling and real-world application by incorporating the model into a comprehensive framework for mosquito abatement planning, as discussed in section \ref{Evaluating_spraying_strategies}.
Integrating real-time weather data and patchy environment have allowed us for a dynamic, adaptive planning approach that has optimised interventions based on current and forecasted conditions. 
Through simulations under various abatement strategies and weather conditions, our current study provides a practical tool to aid the public health authorities and mosquito control districts more effectively curb mosquito-borne disease transmission. \par

The simulations from our ODE-based model have highlighted that local interventions do not act in isolation but are modulated by the spatial context. 
Simulations from our model have demonstrated that larvicide or adulticide applied in one patch can potentially lead to spillover effects, either reducing the effect of recolonisation from neighbouring patches or being offset by immigration from the untreated sites. 
Such findings from our model have also emphasised the limitations of uniform, one-size-fits-all, homogenous control strategies and the need for coordinated interventions across connected habitats and mosquito abatement districts.\par

The model also has highlighted the complex temporal dynamics of intervention effectiveness. 
While the applications of ULV spray has yielded rapid reductions in adult \textit{Culex} mosquito abundance, the benefits are often short-lived. 
This could potentially be attributed to the recruitment from aquatic stages and migration from untreated-treated patches, whereas larvicide interventions act in a relative slower pace but can provide longer-term suppression by reducing recruitment into the adult stage. 
The combined (both the ULV spray and larvicide) interventions demonstrated synergistic effects, thus suggesting that an integrated control strategy can potentially outperform single interventions.\par

Sobol’s sensitivity analyses further have indicated that parameters associated with mosquito reproduction, natural mortality, and migration rates strongly influence outcomes. 
We have also illustrated the differences between various model parameters in the single-patch and two-patch models, which clearly demonstrate the influence of mosquito dispersal and the consequent reduction in the overall effectiveness of ULV applications in a multipatch environment compared to a single-patch setting.
This can potentially imply that accurate local data on mosquito dispersal and site-specific mosquito lifecycle traits are crucial for tailoring control strategies to specific landscapes. \par

%\textbf{\textcolor{red}{Have to include: early season spray in a regular manner can potentially reduce the relative abundance and adulticide+larvicide gives the best results}}
It is interesting to notice that in two-patch weather-driven \textit{Culex} population model \eqref{Eq3}, our simulations suggest that a well-coordinated ULV spraying campaign implemented early in the mosquito season can potentially yield relatively higher effective outcomes compared to those interventions conducted during the peak of summer as described in the Table \ref{table:A}.
An early-season ULV spray can act as a proactive measure while suppressing the initial rise of adult \textit{Culex} abundance before the favourable temperature conditions accelerate population growth and movements between patches.
This spraying schedule not only reduces the chance of \textit{Culex} population rebound later in the season but also increases the likelihood of long-term efficiency of abatement programmes.
When the ULV spray applied periodically and in synchrony across the connected habitats, such strategies could enhance the chance of a sustained suppression of \textit{Culex} abundance, thus presenting a more resilient and cost-effective approach for vector control.
We also observe that the coordinated application of ULV adulticide and larvicide in a two-patch weather-driven model illustrates a significantly greater reduction in \textit{Culex} abundance compared to the use of ULV spraying alone. 
Simultaneous employment of larvicide effectively disrupts the aquatic stages of \textit{Culex}(eggs, larvae and pupae), thereby interrupting the regeneration cycle of \textit{Culex} population while the ULV spraying mainly targets the adult \textit{Culex} mosquitoes.
This combined strategy can potentially provide a more extensive and sustained suppression of \textit{Culex} populations across connected habitats. 
Consequently, inclusion both ULV and larvicide treatments as a regular, synchronised component of mosquito abatement programmes can increase long-term control efficiency and reduce the likelihood of population resurgence following adulticide-only interventions.

%\textbf{\textcolor{red}{Similarities- dissimilarities  with previous work:}}

The results described in the section \ref{Evaluating_spraying_strategies} is notably different from the findings in single-patch model as reported in ~\cite{BHOWMICK2025103163, EZANNO201539}, highlighting that inclusion of spatial connectivity through mosquito movements can fundamentally change the predicted outcomes of different control strategies. 
This two-patch modelling framework can potentially capture the dynamics between treated and untreated areas, where migration can either reintroduce \textit{Culex} into the controlled zones or, provide an opportunity to spread of suppression benefits across neighbouring habitats.
Hence, considering mosquito dispersal is a crucial information for correctly evaluating the efficacy of coordinated interventions and for designing more realistic spatially informed mosquito abatement strategies. 
Consistent with earlier studies, the sensitive parameters identified in our sensitivity analyses closely correspond to those reported in previous studies ~\cite{BHOWMICK2025103163, EZANNO201539, CAILLY20127}. 
These include adult \textit{Culex} mortality, egg-laying rates, fecundity, larva-to-pupa transition rates, and the efficacy of ULV spraying. 
There are some discrepancies due to differences in the response variables considered in ~\cite{EZANNO201539} compared to ours. 
In contrast, the authors in ~\cite{BHOWMICK2025103163} have employed a similar response variable but within a simpler, single-patch modelling framework and the findings are in line with ours.
Overall, our results and the dynamics of our model are broadly in agreement with the findings of ~\cite{BHOWMICK2025103163, EZANNO201539, CAILLY20127, YU201828}.
The mathematical formulation of the basic offspring number in our study is consistent with the expressions that reported in ~\cite{YANG_MACORIS_GALVANI_ANDRIGHETTI_WANDERLEY_2009, BHOWMICK2025103163, doi:10.1142/S0218339015500278}. 
The functional dependence and overall magnitude of ( $R_0$ ) on temperature also exhibit strong similarity to the trends described by ~\cite{BHOWMICK2025103163, 10.7554/eLife.58511}. 
Furthermore, our simulation outcomes support the conclusions of ~\cite{BHOWMICK2025103163, 10.1093/jme/tjae041, 10.1093/jme/tjad088, 10.1093/jme/tjaf024}, indicating that adulticide treatment alone fails to achieve a consistent or long-term suppression in the relative abundance of \textit{Culex} population or in the magnitude of ( $R_{0}$ ) but a combination of adulticide and larvicide can abatement the mosquito population ~\cite{Unlu}. 
The \textit{Culex} trait-based ( $R_0$ ) models developed here effectively has isolated the physiological influence of temperature on fecundity, and the influence of dispersal between the patches.
The resulting functional relationships of ( $R_0$ ) qualitatively agree with those reported by ~\cite{BHOWMICK2025103163, 10.7554/eLife.58511}. 
Additionally, the graphical pattern of ( $R_0$ ) as a function of temperature closely mirrors the characteristic curves described by these authors.

%\textbf{\textcolor{red}{Limitation:}}
Several limitations of our current weather-driven metapopulation model of \textit{Culex} population should be acknowledged. 
We have assumed the homogeneous environmental conditions within each patch to construct the model, whereas real landscapes exhibit variability in microclimate, site specific fecundity or mortality ~\cite{Janet, 10.1093/jme/tjad033, Endo, Felix, Villena,10.1603/0022-2585-41.4.650}.
To address the incomplete information regarding different landscape types and associated factors, we have conducted a series of simulations under varying initial conditions, weather scenarios, and carrying capacities, the details of which are provided in the SI.
We have simplified certain entomological processes, such as temperature-dependent survival, diapause dynamics, and overwintering etc.  to maintain model tractability and this has enabled us to perform the simulations while considering the heterogeneity in the temperature-rainfall data ~\cite{BHOWMICK2025103163, EZANNO201539}.
Alongside, we have not explicitly incorporated stochastic events such as extreme weather events or habitat disturbances or, seasonality, which can markedly influence the \textit{Culex} population trajectories ~\cite{su17010102, PADDE2024107417, 10.3389/finsc.2023.1144072}.
We have utilised the entomological and ecological parameter values that were derived from literature-based estimates and they may not fully able to capture local variability or temporal fluctuations in the \textit{Culex} traits, control efficacy and dispersal strength.
The way we have modelled the control interventions-particularly ULV spray is idealised and we have not included any operational uncertainties such as spray coverage, timing inaccuracies, insecticide resistance or, the variability in the efficacy of ULV and larvicide ~\cite{10.1371/journal.pcbi.1006831, SCHLEIER201272, SAVVIDOU2025112521}.
To account for uncertainty in these factors, we also have performed a detailed sensitivity analysis.
Finally, while we have modelled the relative abundance of the \textit{Culex} population, we have not directly coupled the weather-driven ODE based metapopulation modelling framework to host infection dynamics and therefore, implications for disease transmission remain indirect ~\cite{BHOWMICK2024107346, BHOWMICK2023110213}.
The current framework does not allow us to include of site-specific infection indicators or surveillance metrics that we could have used to refine and adapt spraying regimes in response to ongoing transmission levels.
Although assessing these factors in detail is beyond the scope of our current study, the weather-driven metapopulation model can readily be adapted to include their parameterisation in future work.
\par

%\textbf{\textcolor{red}{Future work:}}
Future research should concentrate on extending this framework in several directions and focus on to address these limitations to strengthen the model’s ecological realism and predictive results while considering the fluctuations in weather data. 
Coupling this modelling framework with high-resolution climate and landscape data, as well as fine scale mosquito dispersal network could facilitate a better understanding of \textit{Culex} population dynamics and potential disease burden in complex landscape ~\cite{10.1093/jme/tjaf140, Krol, Peter}.
Incorporating various spatial heterogeneity in environmental-entomological variables, such as temperature gradients, breeding habitat quality, mosquito dispersal rate, distribution of overwintering \textit{Culex} mosquitoes and availability of blood meal, can enable more accurate representation of \textit{Culex} population dynamics ~\cite{SAARMAN2025101205, Romit, SAUER2023100572, ijerph16081407, 10.1371/journal.pntd.0012245}. 
Integrating adaptive control modules that consider insecticide resistance, operational delays, and varying coverage levels could further strengthen the robustness of intervention simulations ~\cite{Sass-2022-moco-38.1, Noel}.
Additionally, introducing extreme weather events such as heatwave, flash flood, sudden cold snap, heterogeneity in breeding site productivity, or anthropogenic driven changes in the landscape would capture the variability of \textit{Culex} population dynamics ~\cite{w14010031, https://doi.org/10.1111/jvec.12265, IIia, FERRAGUTI2024109194}. 
Finally, advancing this weather-driven metapopulation modelling framework through agent-based or network-driven: integrating mosquito dispersal and ULV spray delivery networks, approaches could support a better predictive model to forecast the relative abundance of \textit{Culex}. 
This can also help to validate under more realistic field conditions and thus contributing to the design of evidence-based and spatially targeted mosquito abatement plannings.

%\textbf{\textcolor{red}{Conclusion:}}
In conclusion, this study illustrates that modelling \textit{Culex} mosquito population in a patchy environment can provide a more realistic understanding of vector dynamics and the potential impacts of various control interventions. 
The results demonstrate that: Spatial heterogeneity matters: mosquito dispersal amongst the patches can undermine the local interventions if the surrounding habitat patches are not being treated. 
Therefore, a well-coordinated synergistic approach is the most appropriate way for the abatement of mosquitoes. 
Intervention type and timing are critical: ULV spray can provide a rapid yet short-lived reduction, while the application of larvicide can deliver slower but more sustained control; a combined application of ULV and larvicide are the most effective. 
Coordination increases effectiveness: The mathematical formulation of the Basic Offspring Number of the metapopulation model has provided a mathematical framework to pave the way for the abatement strategies and they have to be collaborative for the reduction than fragmented approaches.

\bibliography{MosPopoModel}

\end{document}